\documentclass[letterpaper, singlespace, twocolumn]{emulateapj}
\newenvironment{unenumerate}{\def\item[]{\smallbreak\noindent $\bullet$\enspace}}{\smallbreak}

\usepackage{natbib} 
\usepackage{graphicx}
\usepackage{bm}
\usepackage{hyperref}
\usepackage{subfigure}
\usepackage{amsmath}
\usepackage{amssymb}
\usepackage{wasysym}
\usepackage{verbatim}

\def\m2h{$^{1\text{D}}$MESA2HYDRO$^{3\text{D}}$}

\begin{document}

\title[]{Density Conversion between 1-D and 3-D Stellar Models with $^{1\text{D}}$MESA2HYDRO$^{3\text{D}}$}

\author{M. Joyce\altaffilmark{1,2.3}, L. Lairmore\altaffilmark{4}, D. J. Price\altaffilmark{5}, T. Reichardt\altaffilmark{6}  and S. Mohamed\altaffilmark{2,7,8}
}

\affil{$^{1}${Research School of Astronomy and Astrophysics, Australian National University, Canberra, ACT 2611, Australia}}
\affil{$^{2}$Department of Physics and Astronomy, Dartmouth College, Hanover, New Hampshire, 03755, USA}
\affil{$^{3}$South African Astronomical Observatory, Observatory Road, Cape Town, South Africa}
\affil{$^{4}$Robotics Engineering Department, KeyMe, New York, NY, 10005}
\affil{$^{5}${Monash Centre for Astrophysics (MoCA) and School of Physics and Astronomy, Monash University, Vic. 3800, Australia}}
\affil{$^{6}${Department of Physics and Astronomy, Macquarie University, Sydney, NSW 2109, Australia}}
\affil{$^{7}$Department of Astronomy, University of Cape Town, Private Bag X3, Rondebosch 7701, South Africa}
\affil{$^{8}$South African National Institute for Theoretical Physics, Private Bag X1, 7602 Matieland, South Africa}

\begin{abstract}
We present \m2h, an open source, Python-based software tool that provides an accessible means of generating physically motivated initial conditions (ICs) for hydrodynamical simulations from 1-D stellar structure models.

We test \m2h on five stellar models generated with the MESA stellar evolution code {and verify its capacity as an IC generator with the Phantom smoothed-particle hydrodynamics code \citep{MESAIV, Phantom}.
Consistency between the input density profiles, the \m2h-rendered particle distributions, and the state of the distributions after evolution over $10$ dynamical timescales is found for model stars ranging in structure and density from a radially extended supergiant to a white dwarf.
}

\end{abstract}

\maketitle

\section{Introduction}

The objective of stellar modeling is to reproduce the observed characteristics of stars with as much fidelity {to physical reality} as possible. Given unlimited time and computational resources, we would simulate entire stellar life cycles in rigorous, three-dimensional detail;
however, it is not currently computationally feasible to compute the evolution of a star {particle-by-particle} in 3\nobreakdash-D. 
{Because of this practical limitation,} one-dimensional stellar structure and evolution codes (SSECs) are used {to model the secular evolution of stars over astronomical timescales (e.g.,~DSEP, Y$^2$, BaSTI, MESA, and many others \citealt{Dotter, YY, VDB2006, Piet04, BaSTI, MESAIV}). Models of this type are sufficient and highly effective for a wide range of applications, but they are not well-suited to some important, intrinsically three-dimensional phenomena.} Processes such as dynamical stellar behavior, stellar interactions, mergers, and convection cannot be captured realistically {using 1\nobreakdash-D codes. Rather, astrophysical problems of this nature are best approached using full 3-D modeling for isolated scenarios or on shorter physical timescales.
The appropriate tool for such problems is a 3-D hydrodynamics code \citep{GADGET, PriceSPH, Mohamed12, GIZMO, BoothMohamed16, RamstedtMohamed17,Goldbaum, Phantom}.
}

With the many SSECs now {in use}, astronomers can produce models that take into account a wide range of physical processes, including elemental diffusion, gravitational settling, fine-tuning of mixing events, and elaborate meshes for specifying the initial abundances of over 50 elements individually. The highly customizable nature of these codes makes them especially powerful, as the user may designate any number of physical attributes (mass, abundances, equation of state, opacity tables, mixing prescriptions) and recover the surface or structural state of the model star at any point in evolutionary time.

To make them as realistic as possible, SSECs integrate externally generated grid data, including opacities, model atmospheres, and mixing specifications, which are calculated using 3-D simulations or other more sophisticated methods (\citealt{Ig, Ferg, Hauschildt, Trampedach07, MagicMLT, 321D}). 
Fundamentally, however, SSECs are still based on a one-dimensional formalism, and as such are not capable of capturing {particle-driven} processes themselves. 
Rather, the {modeling techniques} best suited to these tasks are those that compute the properties of large numbers of fluid elements over short timescales in three dimensions; {these include hydro\-dynamics codes such as GADGET-2, AREPO, AstroBEAR, Phantom, and many others (\citealt{GADGET, AREPO, GIZMO, AstroBEAR, Phantom}). } 
Thorough reviews of the hydrodynamical modeling landscape {and summaries of the capabilities and mechanics of many such codes are provided in, e.g.,~\citet{Rosswog,Springel2010, PriceSPH, Goldbaum}. }

A critical aspect of hydrodynamical modeling is the specification of the {initial distributions in mass, energy, and other physical quantities} 
at the onset of the simulation. {Often, the initial mass distribution} that represents a star, or stellar atmosphere, is {obtained} by generating an arbitrary particle distribution that follows a $1/r$ profile \citep{PriceSPH}. 
{While sometimes appropriate and sufficient, this choice} is often made ad hoc and may not reflect the actual distribution of material in a star {very accurately}.

The need for and feasibility of {other} techniques to construct initial conditions {with more physical fidelity} has been recognized, for instance, in the work of \citet{Pakmor2012}, who showed that a method for building a 3\nobreakdash-D configuration using concentric, shellular distributions of particles was effective for modeling white dwarfs.
More recently, \citet{Ohlmann} demonstrated that a 1-D density profile generated directly with MESA (Modules for Experiments in Stellar Astrophysics; \citealt{MESAIV}) could be used to construct initial conditions for 3-D hydrodynamical models of red giants. This technique has inspired the development of {our package.}

\m2h is an open source initial conditions (IC) generator for hydrodynamical simulations that use equal-mass particles. 
Generalizing and extending the work of \citet{Ohlmann}, \m2h constructs a particle distribution based directly on an SSEC-generated, radial density profile provided by the user.
Because it samples a stellar structure model directly, \m2h makes it easier to incorporate the customizable physics of 1-D SSECs into 3-D hydro\-dynamical simulations.

In this paper, we present the release of \m2h, describe our methods, and demonstrate consistency between MESA density profiles and the particle distributions \m2h generates from them. 
{We test our package on five model stars with diverse structures encompassing a density range of $18$ dex. We demonstrate that the tool can effectively render density profiles as computationally tractable, physically adherent particle distributions across a broad span of parameter combinations. We directly verify \m2h's utility as an IC generator using the Phantom smoothed-particle hydrodynamics code.}

\section{MESA Models as Initial Conditions}

\m2h can convert any radial density profile {that is smooth, physically reasonable, and formatted compatibly.} However, we have focused our efforts towards integration with the MESA code for a few reasons: (1) the MESA community is large, spans a broad range of interests, and adheres to an open-source ethos, (2) the MESA code is under active development, with new releases every few months, and (3) MESA has the highest degree of customizability of any stellar evolution code.
{The user can control thousands of parameters,} many of which impact the structural aspects of the model star. The preservation of such effects is the primary benefit \m2h provides over less complicated methods for {prescribing density profiles} in hydrodynamical initial conditions.

MESA is a suite of open source, thread-safe libraries developed in Fortran 95. It contains distinct modules for handling the equations of state, opacity, nuclear reaction rates, diffusion data, and atmospheric boundary conditions. Each module is a separate library with its own public interface, supports shared memory parallelism based on the OpenMP application program interface, and employs adaptive mesh refinement and time step control. 
Extensive testing of MESA indicates that the code effectively calculates the evolution of stars over a wide range of masses (from brown dwarfs to $M=90\, M_{\odot}$), and over evolutionary phases spanning from the pre-main sequence to the onset of core collapse in high-mass stars, or to the white dwarf stage in lower-mass stars. Detailed information on the workings of MESA is available in their several instrument papers, including \citet{MESAI, MESAII, MESAIII,MESAIV}.

The physical attributes of a stellar model are specified in MESA via a central control file called an ``inlist.'' This contains {all modeling, physical, and plotting controls used in the evolutionary run.}
The public release of \m2h includes copies of {the inlists} used to generate our test stars, where variable names {have the same definitions as in MESA version 10398}.

{Throughout the course of a run, MESA generates ``profile'' files, or snapshots of the state of the star from core to surface, as a function of radius (equivalently, mass contained). The user can control the output frequency of these snapshots.
Meanwhile, a ``history'' file tracks the secular evolution of the global state variables---mass, radius, effective temperature, average density, etc., and the quantities derived from these---as a function of time.} 
{One can thus recover the structure} during any evolutionary phase by specifying stop conditions which correspond to the desired {physical criteria}.
To obtain a structural model of a thermally pulsing asymptotic giant branch (TP-AGB) star, for example, the user can simply specify that MESA terminate evolution during a thermal pulse (e.g., \citealt{TUMi}).

\m2h requires that the input MESA profiles have a particular data organization. 
{For reference, we include the complete glossaries of MESA parameter defaults (as they appeared in MESA version 10398) as well as the particular settings used to generate our test suite for each of the four standard MESA control lists: ``controls,'' ``star\_job,'' ``history\_columns,'' and ``profile\_columns.'' The latter-most of these dictates the data format that should be used when generating profiles intended for use with our package. 
The user may also provide a stellar structure model generated by another SSEC as long as it is in a format understood by \m2h.   
For instance, the data columns in a compatible model must adhere to the MESA naming conventions, using e.g., ``logR'' as the column name for ($\log_{10}$) radius.
The \m2h User's Guide gives an example of a correctly formatted input profile and provides the column header keywords (``logR,'' ``mass,'' ``logRho,'' etc.) needed to build one's own input model, if desired. The manual also provides more detail on MESA control files. It is freely available by request from the authors prior to public release. }

{The physical quantities required by \m2h are, minimally, mass, density, pressure, and internal energy as a function of radius---the latter two of which are necessary to ensure piecewise compatibility between the equations of state in the 1-D and 3-D models and to compute smoothing lengths for the SPH particles.}

{MESA inlists corresponding to the five stars tested in this paper are included with the package. These serve as a useful starting point, but the user should change parameters to suit their needs. Some knowledge of the appropriate astrophysical choices is required in order to generate reasonable models.}

\section{Methods}
\label{section:methods}
The {backbone of} the 1-D to 3-D mapping is the conversion of discrete $r$,~$ \rho(r)$ data to a set of {mass and radial coordinates that can be reinterpreted as particle distributions. These are referred to henceforth as} $N,R$ coordinates, abbreviated for ``number'' and ``radius.''

{The radial coordinate $R$ represents the value at which a particular shell must be located, relative to the stellar core, in order to recast the mass contained in the region $ (r_u - r_l)$ equivalently as a distribution of discrete particles covering the surface of a sphere.
The $R$ coordinates are actually midpoint values $ r_{\text{mid}} = (r_l + r_u) / 2 $, where $r_l$, $r_u$ are the inner and outer bounds, respectively, on the mass region to be mapped. The number $N$ corresponds to an integer used by the HEALPix spherical tessellation algorithm to dictate the total number of particles $n_p$ distributed over a single spherical shell \citep{HEALPix}. This coordinate, in conjunction with the user's choice for particle mass $m_p$, controls mass.

For all models in the current test suite, the same value of $N$ is used for every shell; in a sense, this makes the choice of $N$ a global parameter.
However, it is prudent to treat $N$ as a local coordinate because it is set independently per shell and the value of $R$ depends on it. As such, it is possible to create particle distributions with varying $N$ or to stitch together coordinate files parameterizing different regions of the same model using distinct $N$ values, though we do not do so here.}

The calculation of the $R$ coordinates {relies on a root finding procedure which searches the 1-D profile data passed initially; as such, \m2h works best for profiles which are relatively smooth. Though the algorithm can handle density profiles that are nonmonotonic, it is slower in these regions.} 
The number of $N,R$ coordinates generated {for a given density profile} corresponds to the number of shells used in the 3-D distribution. Shell counts on the order of a few hundred to a few thousand will produce distributions comprising {particle numbers that are} manageable for most hydro\-dynamics codes.

For a hydrodynamical simulation to be {physically meaningful}, it is important that the initial conditions do not produce preferred directionality or organized motion when the simulation evolves \citep{ShazreneThesis}.
One method for building stellar particle distributions with the appropriate degrees of randomness and point separation is to cut a spherical region out of a ``glass,'' or periodic box, which is randomly populated with particles.
Another way is to stack individual shells of particles concentrically, building up the star layer by layer.

\citet{Pakmor2012} couple the stacking technique with surface distributions generated via the Hierarchical Equal Area iso-Latitude Pixelization, or HEALPix, tessellation algorithm \citep{HEALPix}. HEALPix works by subdividing the surface of a sphere into $n_p$ quadrilateral regions, or cells, of equal area and placing exactly one particle in each cell. {The strength of this} distribution method is that it provides both smooth and random initial conditions, and thus minimizes the emergence of non-physical artifacts during the system's evolution \citep{ShazreneThesis}.
{Another benefit to this technique is that HEALPix uses only the desired number of particles, or, equivalently, cells, to generate the appropriate set of $(x,y,z)$ particle coordinates.} However, the tessellation requires that particle numbers $n_p$ satisfy the condition $n_p = 12 N^2$, with $N=2^x$ for some integer~$x$, {thus considerably limiting flexibility in particle numbers.}

Despite this limitation, we replicate \citet{Pakmor2012}'s concentric stacking method and use of the HEALPix algorithm to construct shellular, surface particle distributions in \m2h. HEALPix is integrated with our package via {healpy} \citep{healpy}, a Python interface to the algorithm that installs automatically with \m2h.

\begin{figure} 
\centering
\includegraphics[width=\linewidth]{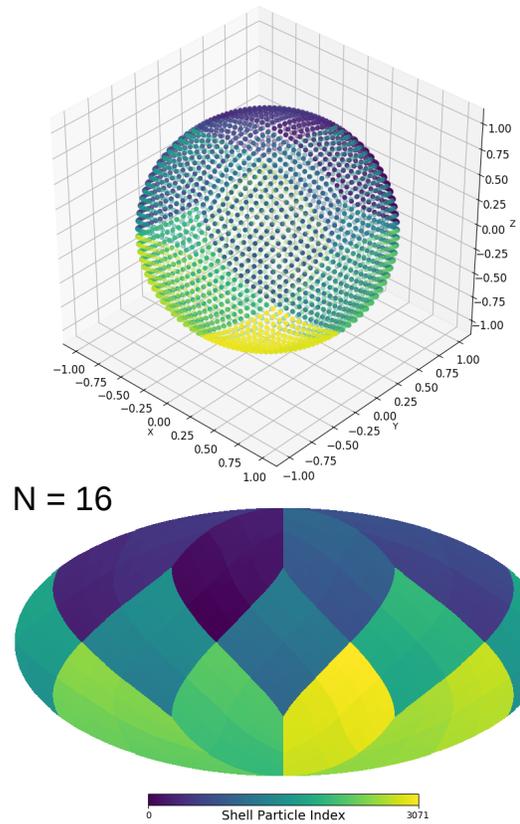}
\caption[HEALPix tessellation and sample particle shell]{Top: A sample particle distribution given by HEALPix for an arbitrary shell~$k$. This shows 3072 particles, corresponding to $N=16$. Bottom: Hierarchical equal-area iso-latitude pixelization of a spherical surface for $N=16$, generated with {\it healpy}. Particles in the upper panel are color-coded according to their host regions, in the lower panel. The color bar represents particle index, which is an integer assigned to each region/particle.}
\label{kshell}
\label{healpix}
\end{figure}

Figure \ref{healpix} shows a shellular particle distribution and the tessellation from which it is derived for $n_p$ = 3072 (equivalently, $N = 16$).
All regions in the lower panel have equal area and host exactly one particle each. Particles in the upper panel are coded so that each particle has the same color as its host region below, where particles and regions are connected by the assignment of an integer 1 to 3072 (i.e., cell 1, in bright yellow in the lower panel, hosts particle 1, bright yellow in the upper panel, and so forth).

Both \citet{Pakmor2012} and \citet{Ohlmann} use stacking methods to build {their models, but the former construct density distributions} which place single SPH particles in equal--volume cells. This geometric constraint requires that every particle exist in a space of size $(r_u-r_l)^3$, {meaning that \citet{Pakmor2012}'s placement radii are determined by the equality of surface cell size and shell separation---two independent} functions of the width $r_u - r_l$. 
While this equal-volume approach gives good uniformity and convergence, it allows little freedom in specifying the locations of shells relative to each other. {While this is perfectly appropriate when modeling high-density stars such as white dwarfs, it does not generalize easily to other cases. Hence, we avoid using this cubical volume constraint.}

\citet{Ohlmann}'s method, on the other hand, allows for the free variation of shell width. 
It is necessary to let the radial separation of shells vary if one wishes to maintain a constant mass or number of particles per shell. This approach results in less-uniform distributions than {the cubical volume approach}, but it is more flexible and thus more appropriate for a tool which aims to be effective {for a wide variety of model stars.} 
In our method, the relative spacing among HEALPix shells---widths of slices $r_{u,k} - r_{l,k}$---is determined by the bounds on the mass integral $m_{\text{shell}}$.
In the case of fixed particle mass~$m_p$, constant mass translates to a constant number of particles per shell, $n_p=12N^2$.

{As central stellar densities can be several orders of magnitude larger than the local densities in the outer layers of the star, the number of equal-mass particles required to accurately represent the physical density gradients can quickly become intractable. }
\citet{Ohlmann} use only a small percentage of MESA's total density profile in their constructions, electing to build a particle distribution which represents only the outer layers of the star.
{They represent the rest of the interior by a single particle, or ``core,'' whose mass is equal to the mass represented in the distribution subtracted from the total stellar mass. They join these components gravitationally using a modified polytrope}.
While approximate, this technique preserves the region of the star that is involved in dynamical behaviors such as mass transfer and atmospheric pulses, {and it provides an effective alternative to dealing with enormous numbers of particles. Core-cutting also allows one to} capture the majority the star by radius while only dealing with a small fraction of the mass. 

We implement a similar method in \m2h, allowing the user to specify the size of the extracted core either by radius or by mass. {However, we caution an important caveat: While \m2h's core-cutting technique will always produce a particle distribution that sufficiently resembles the initial, 1-D density profile structurally, the situation is more complicated in 3-D.  In order for this to work in a smoothed-particle simulation, the hydrodynamics code must be able to deal intelligently with interactions between the core mass and the rest of the particles. For this reason, \m2h classifies the core mass as a ``sink'' particle distinct from the rest when writing hydrodynamic ICs.}

We use the terms {``depth'' or ``depth cut'' and parameters $m_{\text{depth}}, r_{\text{depth}}$} to refer to the percentage of the star in total {radius or mass, respectively}, to which the density profile is fit. For example, {rendering a model with $m_{\text{depth}} =5$\% means} fitting the radial extent of the star which contains the outermost 5\% of the mass.
We note that these depths will often correspond to regions much deeper than the physical definition of a stellar atmosphere, but we sometimes use the term ``atmosphere'' in this context to emphasize {that penetration percentages are defined with respect to the stellar surface.
{The upper panel of Figure \ref{push_ru} highlights a region corresponding to a 5\% depth cut, by mass, in the context of a MESA density profile representing an AGB star.}

\citet{Ohlmann} use atmospheric mass depths ranging from 1\% to 10\% in their models. To compare our models self-consistently, we fix the penetration depth to a constant percentage by radius, $r_{\text{depth}}=0.75M_{\star}$, in our 1-D validation. This corresponds to mass depths between $0.5$ and $7$\% for all of our model stars.

\section{Algorithm}

{\m2h reads input data and run time parameters from text files organized into \verb|keyword, value| pairs. These are called ``configuration files'' and use a ``.cfg'' extension. The format of a configuration file follows the convention of a Fortran namelist, as is commonly used in other stellar modeling packages.}

The algorithm for computing a set of shell radii proceeds as follows:
\begin{unenumerate}
	\item[] A discrete density profile $r,\rho(r)$ corresponding to {some percentage of the mass or radial distribution (as specified by the user)} is extracted from a smooth MESA profile or similarly formatted file. {The region representing the core mass is separated from the region to be rendered as SPH particles. These sections are shown in black and blue, respectively, in the top panel of Figure \ref{push_ru}. }
	
	\item[] 
	{At the base of the atmosphere, we search for a solution to the equality
	\begin{equation}
	m_{\text{shell}} = \int^{r_u}_{r_l} 4\pi r^2 \rho(r)\, dr = (12 N^2)m_p,
	\label{int}
	\end{equation}
	by shifting the upper bound on the mass shell integral, $r_{u,0}$, surface-ward from its local position until the integrated mass and mass from the summation of HEALPix particles are equal to within some user-defined tolerance, $\delta_\mathrm{TOL}$. 
	{The bottom panel of} Figure \ref{push_ru} demonstrates the selection of an upper radius.

	In equation \eqref{int}, $m_p$ is the mass per particle and $N$ is the HEALPix integer, both of which are set by the user. The choice of $m_p$ and $\delta_\mathrm{TOL}$ have the largest effect on the computation time: higher values of $m_p$ translate to less frequent solutions and hence a lower resolution profile and shorter computation times, whereas lower values of $\delta_\mathrm{TOL}$ correspond to increased precision on the location of $r_u$ and thus longer computation times.  
	The default tolerance is $\delta_\mathrm{TOL}=0.01$. The parameter combinations used for our test models are provided in Table \ref{recovery}.

\item[] When one instance of equality \eqref{int} is satisfied, the coordinates $N$ and $r_{\text{mid}} = (r_u + r_l) / 2 $ are recorded in a standard text file with the prefix ``NR.'' For other physical quantities, such as internal energy $E$ or temperature $\log T$, \m2h searches the MESA data directly for the $r$ values bordering $r_{\mathrm{mid}}$ and linearly interpolates between them to produce approximate values for $E(r_{\text {mid}})$, $\log T(r_{\text {mid}})$, etc., as desired. 

	Following the computation of one such $r_u$, the subsequent lower bound $r_{l,1}$ is set to $r_{u,0}$, and the process repeats until $r_{l,1}, r_{u,1}$ again satisfy equation~\eqref{int}.

	\item[] The calculation of placement radii $r_{\text{mid}}$ continues until \m2h has subdivided the profile {into $k$ regions of variable size $(r_u - r_l)_{j}$, where $j = 1,...,k$. Each region $j$} is then uniquely characterized by its $N,R$ coordinate pair. The generation of an NR file can take anywhere from several minutes to several hours depending on the choice of $m_p,\delta_\mathrm{TOL}$, and the penetration depth. 
	The completed NR file is then passed to HEALPix via {healpy}.
	}

	\item[] For each shell $k$, HEALPix distributes $n_{p,k}$ particles across the surface of a sphere with radius $R_k=r_{\text{mid},k}$ using the equal cell method described in Section \ref{section:methods}. 
	{The conversion between an NR file and an SPH-compatible initial conditions file takes only a few seconds for particle numbers less than $10^6$. }

	\item[] {Having obtained $12N^2$ sets of $(x,y,z)$ coordinates} for the associated particles, the shells are stacked concentrically to form a 3-dimensional, hollowed sphere by normalizing each HEALPix shell by its placement radius relative to the total stellar radius.

	\item[] Each shell is arbitrarily rotated with respect to its neighbors in order to avoid ordered particle alignments. The rotated coordinates $(x',y',z')_k$ are computed via the multiplication of $(x,y,z)_k$ by the unit matrices
	{\footnotesize
	\[
	\begin{bmatrix}
	    x'  \\
	    y'   \\
	    z'  
	\end{bmatrix}
	=
	\begin{bmatrix}
		1 & 0 & \phantom{-}0 \\
	    0 & \cos \theta & -\!\sin \theta \\
		0 & \sin \theta & \phantom{-}\cos \theta
	\end{bmatrix}
		\!\!
	\begin{bmatrix}
		\phantom{-}\cos \phi & 0 & \sin \phi \\
		\phantom{-}0 & 1 & 0 \\
	    -\!\sin \phi & 0 & \cos \phi
	\end{bmatrix}
		\!\!
	\begin{bmatrix}
	    \cos \psi & -\!\sin \psi & 0 \\
		\sin \psi & \phantom{-}\cos \psi & 0 \\
		0 & \phantom{-}0 & 1
	\end{bmatrix}
		\!\!
	\begin{bmatrix}
	    x  \\
	    y   \\
	    z  
	\end{bmatrix}
	\]
	}%
	with $\theta$, $ \phi$ and $\psi$ pseudo-randomly generated over the interval $[0,2\pi]$. The pseudo-random number generator used is Python's \textbf{random.random}, a wrapper for the Mersenne Twister algorithm \citep{MersenneTwister}, which uses the current timestamp as a seed. New values of $\theta$, $ \phi$ and $\psi$ are computed for every $k$ {but not for every particle; otherwise, the distribution provided by HEALPix would not be preserved.}

	{
	\item[] The final set of $k$ stacked, rotated, concentric sets of $(x,y,z)$ coordinates and the MESA attributes assigned to them particle-by-particle are output to a file with the prefix ``IC.'' 
	These arrays can be passed to subroutines that organize the data into file structures compatible with various hydrodynamics codes directly. Currently, \m2h supports output in the GADGET-2 unstructured binary and hdf5 formats, the Phantom binary format, and a simple ASCII text file.
	The user may control the precision with which the numerical data are written to the IC file, as well as the format of that file, using flags in the configuration file. For example, \verb|filetype=phantom_binary| will produce a binary file in the Phantom format.

	\item[] \m2h can reload the 3-D data it has generated  directly and reduce it to a 1-D $r, \rho(r)$ curve using binning parameters specified by the user.
	 \m2h does not use a smoothing kernel or any other SPH features. 
	 To verify compatibility with a particular SPH code, the user should apply the smoothing kernel used in that code to the output distribution, or use a tool such as SPLASH \citep{SPLASH, PriceSPH}.  
	 }

\end{unenumerate}

{The integral in equation \eqref{int} is solved using a fourth-order Runge--Kutta scheme with adaptive step size refinement \citep{RungeKutta}. 
This method was found to be more well-suited to our problem than, for example, {Python's scipy.integrate function},
due to the large variation in radial width that can correspond to a fixed shell mass. 
One can provide an initial guess for the integration step size in the configuration file, though this will be adjusted automatically as necessary depending on the local shape of the density profile and on the particle mass and solution tolerance provided by the user. 
An inappropriate choice in step size may prolong the first few shell calculations, but it will not have a large impact on the computation time. Changes in the solution tolerance, however, scale linearly with computation time.

The \m2h workflow is subdivided into two main procedures: the first translates a 1-D density profile to an NR file by calculating the shell placement radii, and the second translates an NR file to an IC file using the radial spacings and the HEALPix tessellation.  
As the former conversion takes much longer than the latter, the subroutines are written to be executable in isolation. 
Within these subroutines, many other components of the workflow can be isolated by manipulating the appropriate flags in a configuration file.}

\begin{figure} 
\centering
\vskip-12pt
\includegraphics[width=0.85\linewidth]{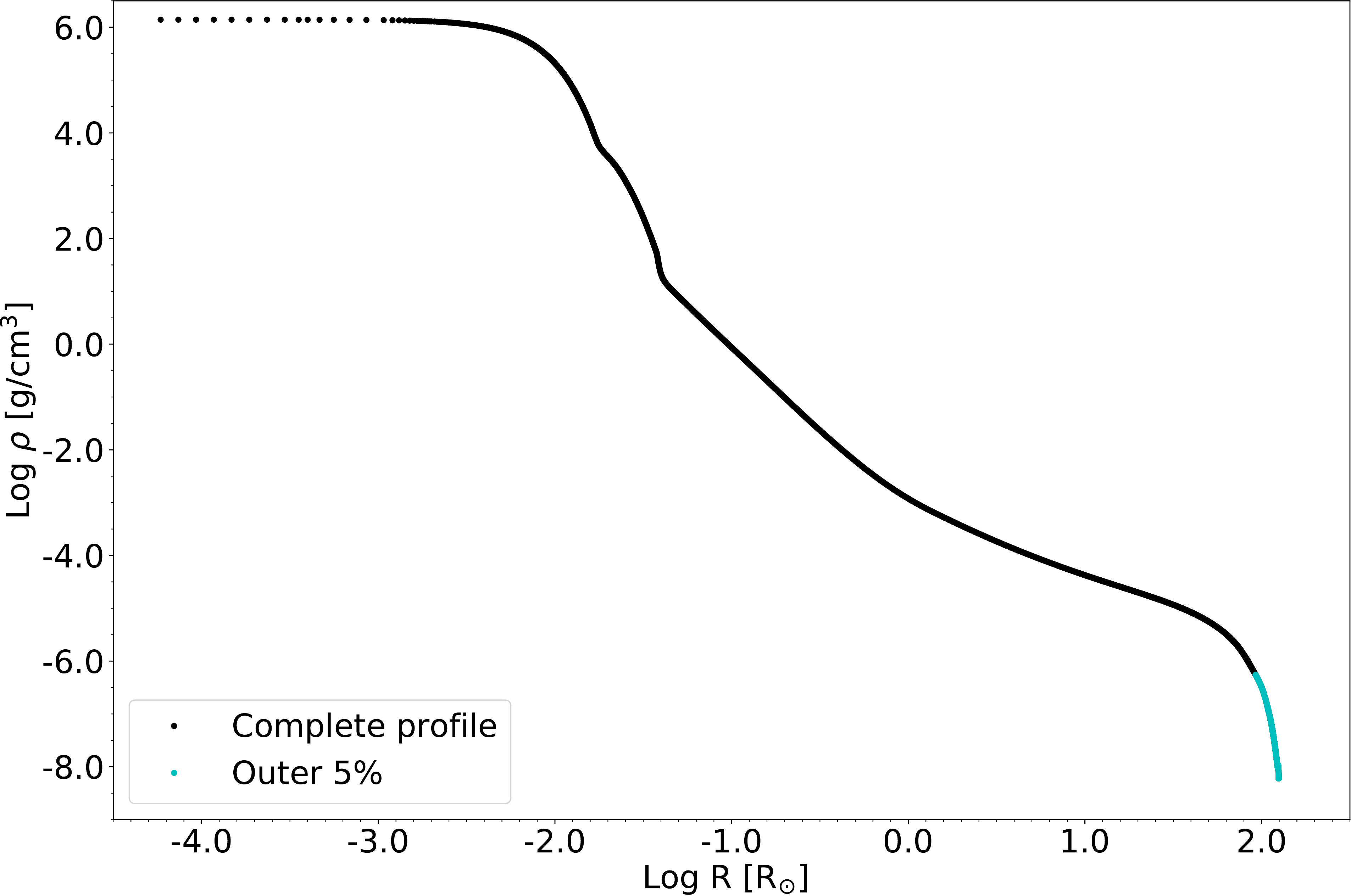}
\vskip5pt
\includegraphics[width=0.85\linewidth]{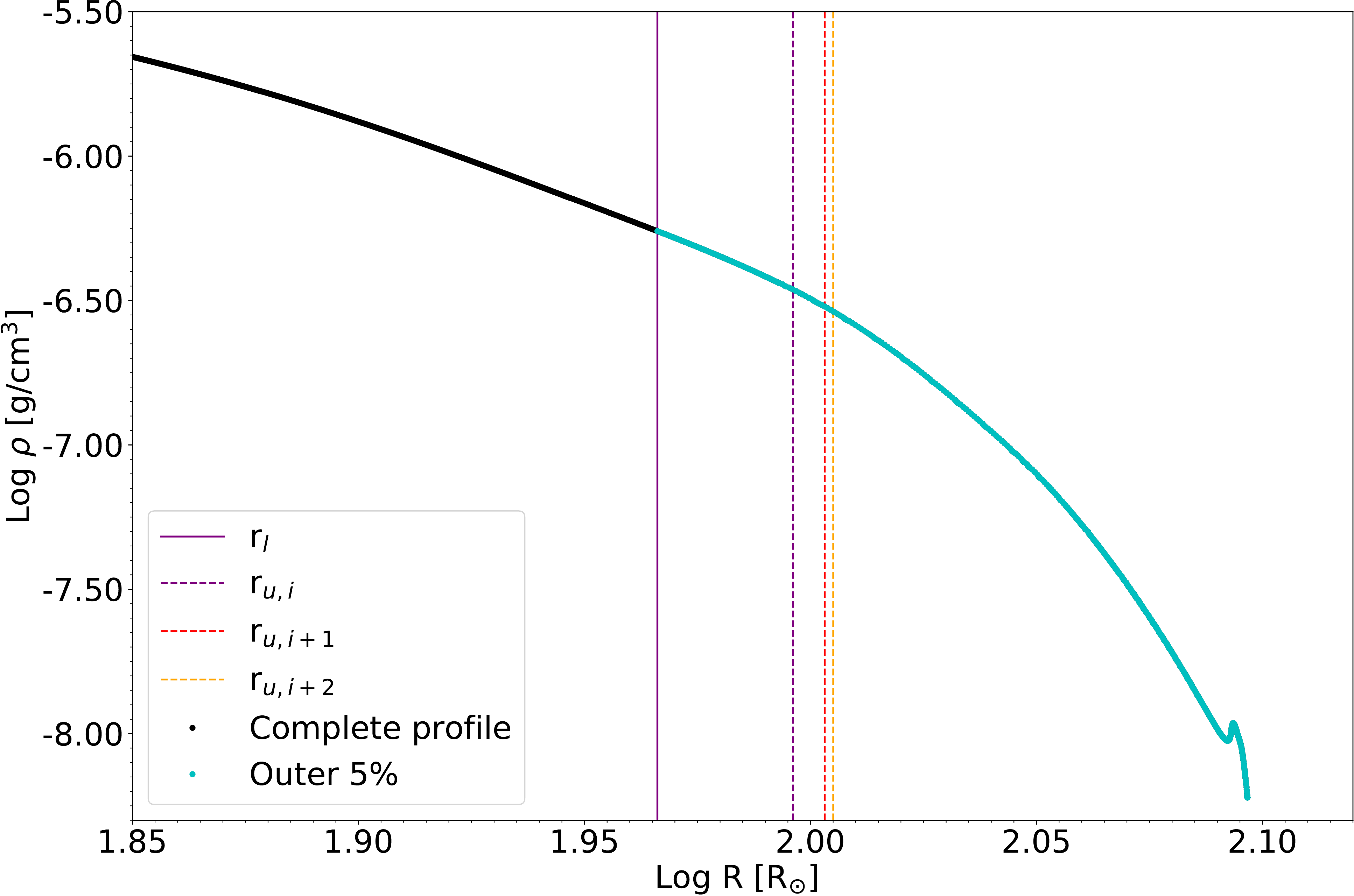}
\caption[Boundary selection for density integral]{%
\small{%
Top: The complete MESA density profile, from core to surface, representing the AGB phase of a $1.8\,M_{\odot}$ model star. The outer 5\% of the star, by mass, is highlighted in blue.
Bottom:  The region we model---isolated in blue in the previous panel.  The left-most vertical line (solid purple) indicates a potential lower integration bound $r_l$, and the three dashed lines to the right indicate trial values of $r_u$, which is pushed rightward, or towards the surface, iteratively until the bounded region satisfies the mass integral in equation \eqref{int}. 
}
}
\label{push_ru}
\end{figure}

\section{Software Package}
\label{package}

\m2h is written in Python but interfaces with various C and Fortran libraries needed by HEALPix, HDF5, {Phantom, SPLASH} and (optionally) the MESA code itself.
{The test suite is verified using data generated with MESA-10398 and Phantom (2019), and so the interfaces are guaranteed to be compatible with these versions of the external software.}

\m2h requires several non-standard Python packages: 
\textbf{
cython
h5py,
hdf5lib,
healpy,
matplotlib.pyplot,
numpy,} and
\textbf{random}.
The main workflow is not currently parallelized, but it is threadsafe so long as the user does not direct the output of two NR calculations to the same file.{\emergencystretch7pt\par}
Our package does not require MESA itself or a hydrodynamics code to run, though it must receive the 1-D profile in a format that uses MESA-style keywords as column headers. 

\m2h can be freely obtained by contacting the authors and installed via the standard Python setup procedure. 
Upon public release, it will be available from GitHub
\begin{verbatim}
https://github.com/mjoyceGR/MESA2HYDRO
\end{verbatim}
or via Python's pip tool.
Detailed installation and setup instructions are provided in the user's guide, also available on GitHub. 

The user can specify ${\sim}20$ parameters in a \m2h configuration files. These include operational attributes (e.g., file names), physical parameters (e.g., particle mass), numerical specifications  (e.g., step size for the numerical integrator), and plotting controls for the recovery test routines (e.g., bin number). 
These parameters may also be specified as command line arguments. Default values are set for all user controls and correspond to a solar-like model star which uses pre-generated MESA data. The program reverts to these values only if neither a configuration file nor any command line argument is specified. A sample configuration file with parameters corresponding to each of the test cases presented here is provided in the \m2h suite.

The operation cycle of \m2h proceeds as shown in Figure \ref{alg}. 
The 1-D input data are processed by routines in the MESA handling library. The data are reduced to a set of $k$ coordinates $N, R$ via methods contained in the numerical routines module, from which HEALPix then generates the requisite set of $(x,y,z)$ coordinates. The complete set of particle positions is written to a file whose format is specified by the user via the input--output (I/O) module. By default, \m2h writes ASCII text files, { but this can result in cumbersome file sizes for large particle numbers. \m2h can produce output in the specific binary formats required by various SPH codes using flags in the configuration files. Some of these} require the HDF5, or Hierarchical Data Format, file format, which is used to organize large data files for supercomputing.
{Fidelity of \m2h's distribution to the initial density profile can then be assessed optionally using the package controls; the user may reload the 3-D distribution, compress it to a 1-D radial curve, and display this against the data loaded from MESA. The most common sources of discrepancy between the input density profile and the output distribution are insufficient tolerance on the integral solution and low resolution imposed by the use of heavy particles.}

A full execution of this cycle may take anywhere from several minutes to a few hours on one thread, depending primarily on the atmospheric depth and the tolerance on equality \eqref{int}. Generating the NR file is the largest source of computation time---typically on the order of minutes to hours. Generating an IC from the NR file does not take longer than several seconds for reasonable particle numbers.  
Exact run times and other computational aspects {for the test suite} are given in Table \ref{recovery}.

{In Figure \ref{alg},} dark blue, square boxes {show components that are the original work of the authors. These can be manipulated directly by the user via the configuration file.} Components represented by green ovals are external software incorporated into \m2h, where interfaces are also the original work of the authors. {Components represented by red ovals are base programs, for example, MESA and Phantom. These can be replaced by any similar SSEC or SPH code, respectively.} Each of these components, as well as all critical subroutines included with \m2h, are described thoroughly in the user's guide.

In addition to the main program, the \m2h package contains a few basic tools. These include a script for confirming a MESA profile is reasonable and a guide for selecting appropriate values for particle mass and initial step size, among others (see the User's Guide for more detail on additional features).

\begin{figure} 
\centering
\includegraphics[width=0.94\linewidth]{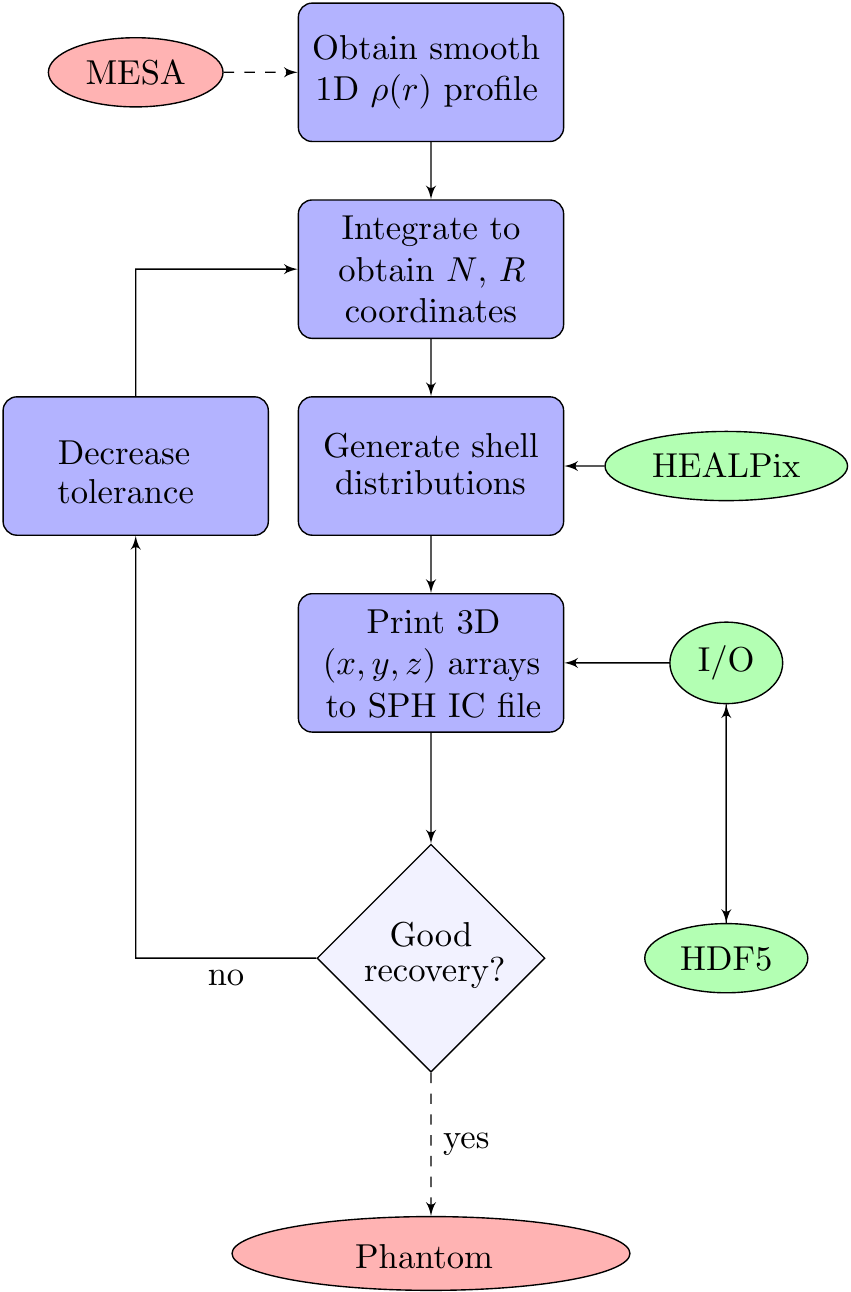}
\caption[Flow of Control Diagram for \m2h]{Flow of control diagram for \m2h.}
\label{alg}
\end{figure}

\section{Validation}

Two situations of particular interest in stellar hydro\-dynamics are (1) pulsation and instability, and (2) dynamical interactions. The latter often involves mass exchange between a radially extended, evolved star and a compact companion, such as a white dwarf. Such stars will necessarily have very different outer density structures, so \m2h must be able to generate initial conditions for stars encompassing a broad spectrum of physical properties.
To demonstrate this capability, we test \m2h on five stellar models which span a range of masses, evolutionary phases, and internal physical configurations. 

We note that the radial extent captured by a given mass depth can vary dramatically depending on the star's density structure; for example, the outermost 30\% of a star by radius contains 5\% of the mass of a $1\,M_{\odot}$, main sequence star, but less than $0.1$\% of the mass of a $90\,M_{\odot}$ supergiant.
{To compare our results systematically, the radial depth is fixed across the set of sample models presented for demonstration, and hence the depth by mass varies among them. }

{Smooth, physically realistic density profiles are produced with the MESA inlists included in this package.  Unless otherwise specified, the 1-D models are generated using the MESA equation of state, the \citet{GS98} opacities, and the ``basic.net'' nuclear reaction network.
Since metallicity is known to impact stellar atmospheres \citep[][and many others]{Thoul03, PortoDeMello08, Asplund09}; we also maintain approximately solar abundance ($Z=0.02$) in each star.
Choices for additional parameters (convective overshoot values, additional nuclear reaction rates, etc.) can be found in the inlists directly. Our choices for the fundamental astrophysical attributes of each test model are discussed below.}


\begin{figure}
\centering
\includegraphics[width=\linewidth]{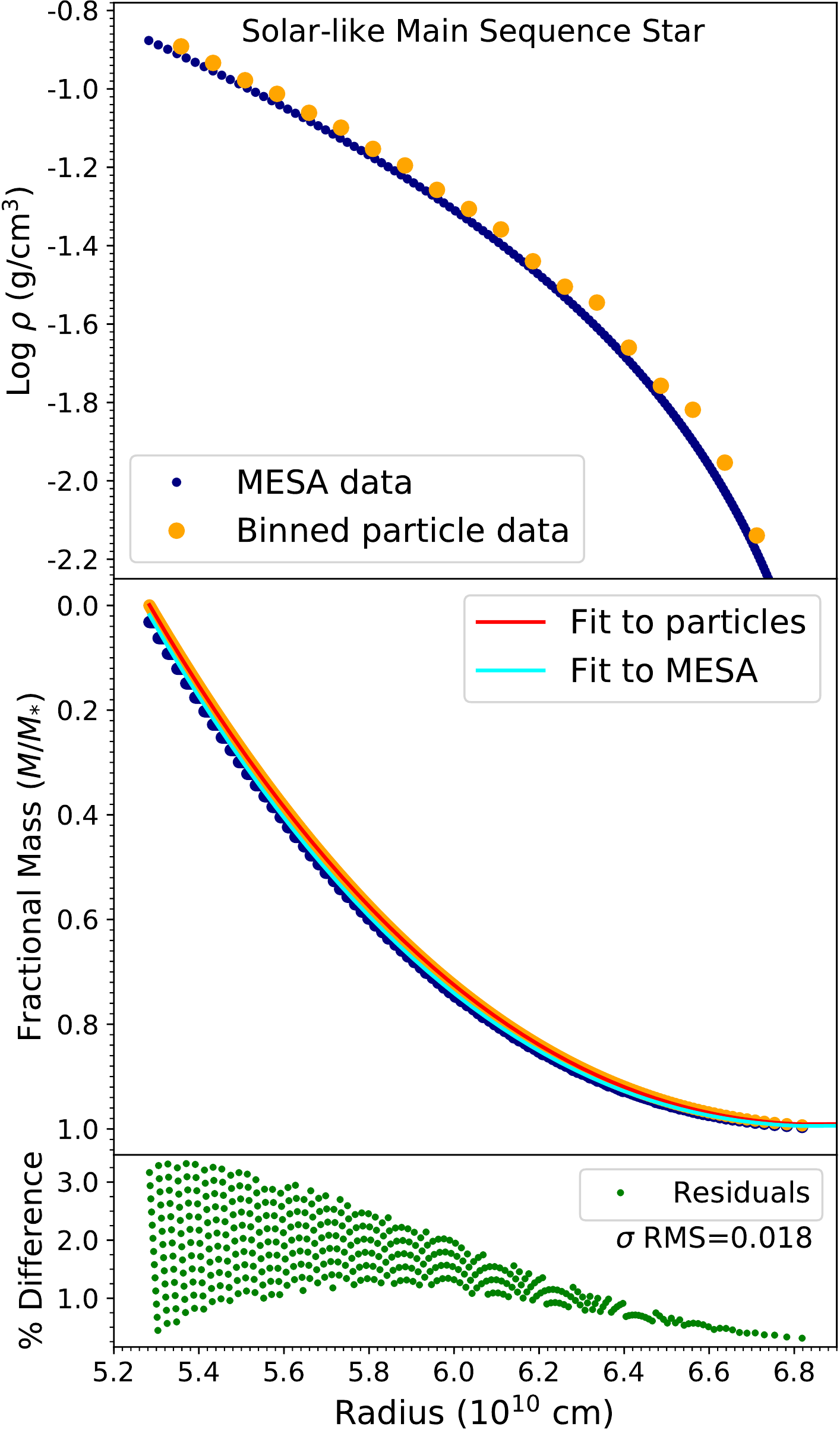}
\caption[Recovery statistics for solar-like main sequence model]{\small 
This panel shows (top) the physical density profile $\rho(r)$ pulled from the input stellar structure model against an approximate 1-D density reconstruction from the reloaded particle data, (middle) the inverted cumulative mass distributions for the input and output data, and (bottom) the residuals between fits to the cumulative mass distributions, as a function of radius, for a solar-like model.
}
\label{starfig1a}
\end{figure}

\begin{figure}
\centering
\includegraphics[width=\linewidth]{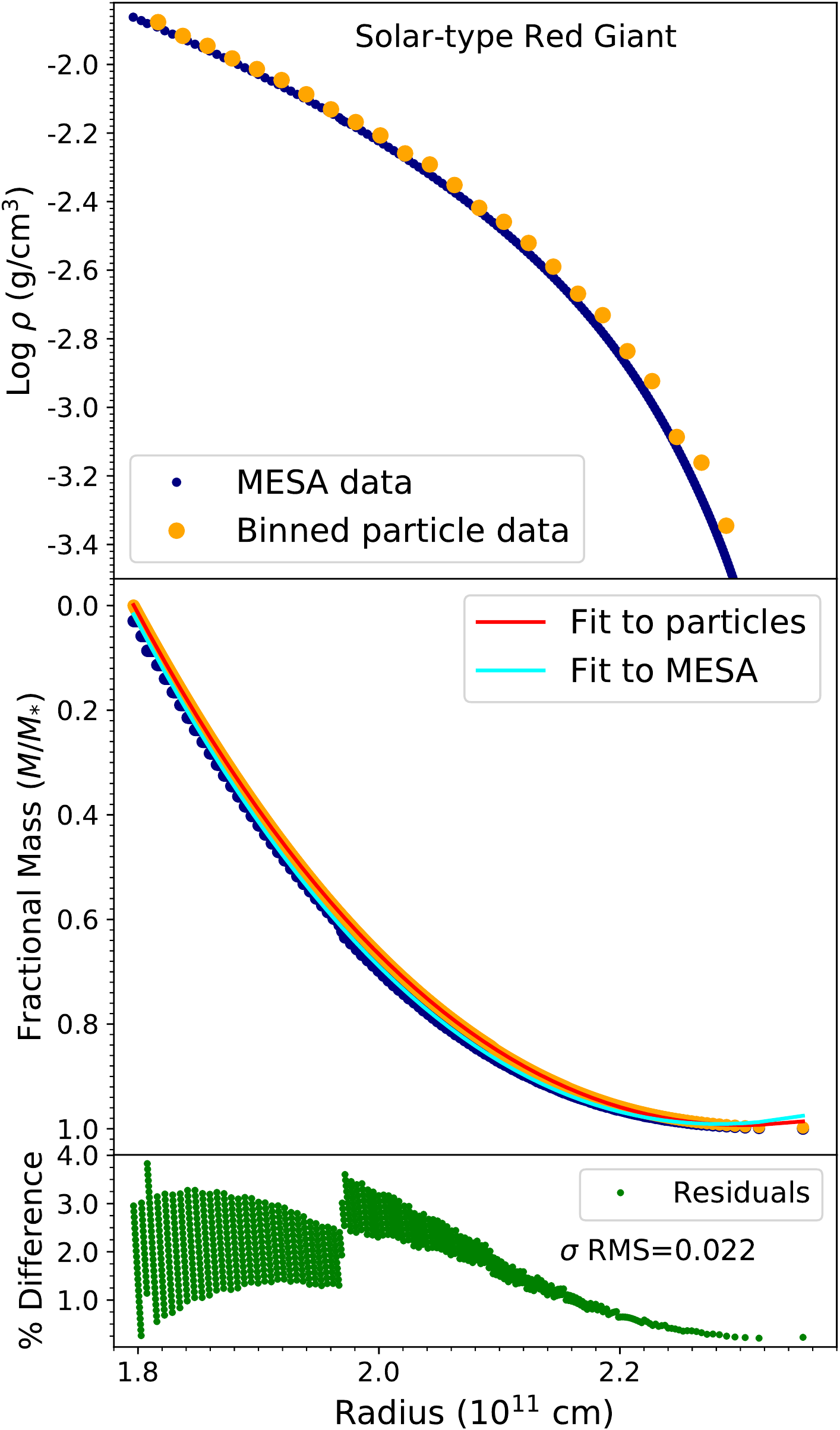}
\caption[Recovery statistics for solar-like red giant]{Same as Figure \ref{starfig1a}, but for a solar-type red giant model.}
\label{starfig1b}
\end{figure}

\begin{figure}
\centering
\includegraphics[width=\linewidth]{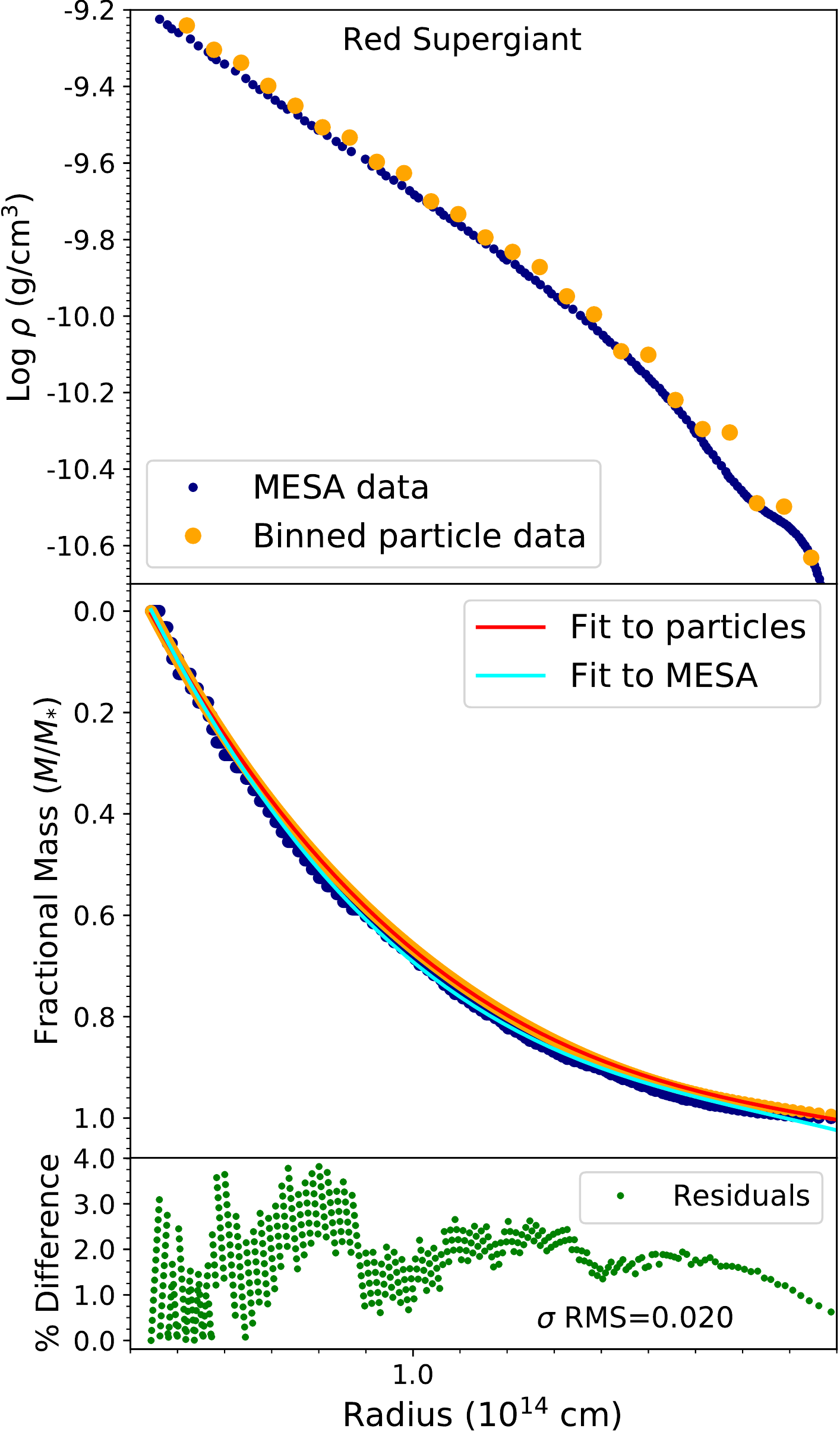}
\caption[Recovery statistics for red supergiant]{Same as Figure \ref{starfig1a}, but for a red supergiant.}
\label{starfig2a}
\end{figure}

\begin{figure}
\centering
\includegraphics[width=\linewidth]{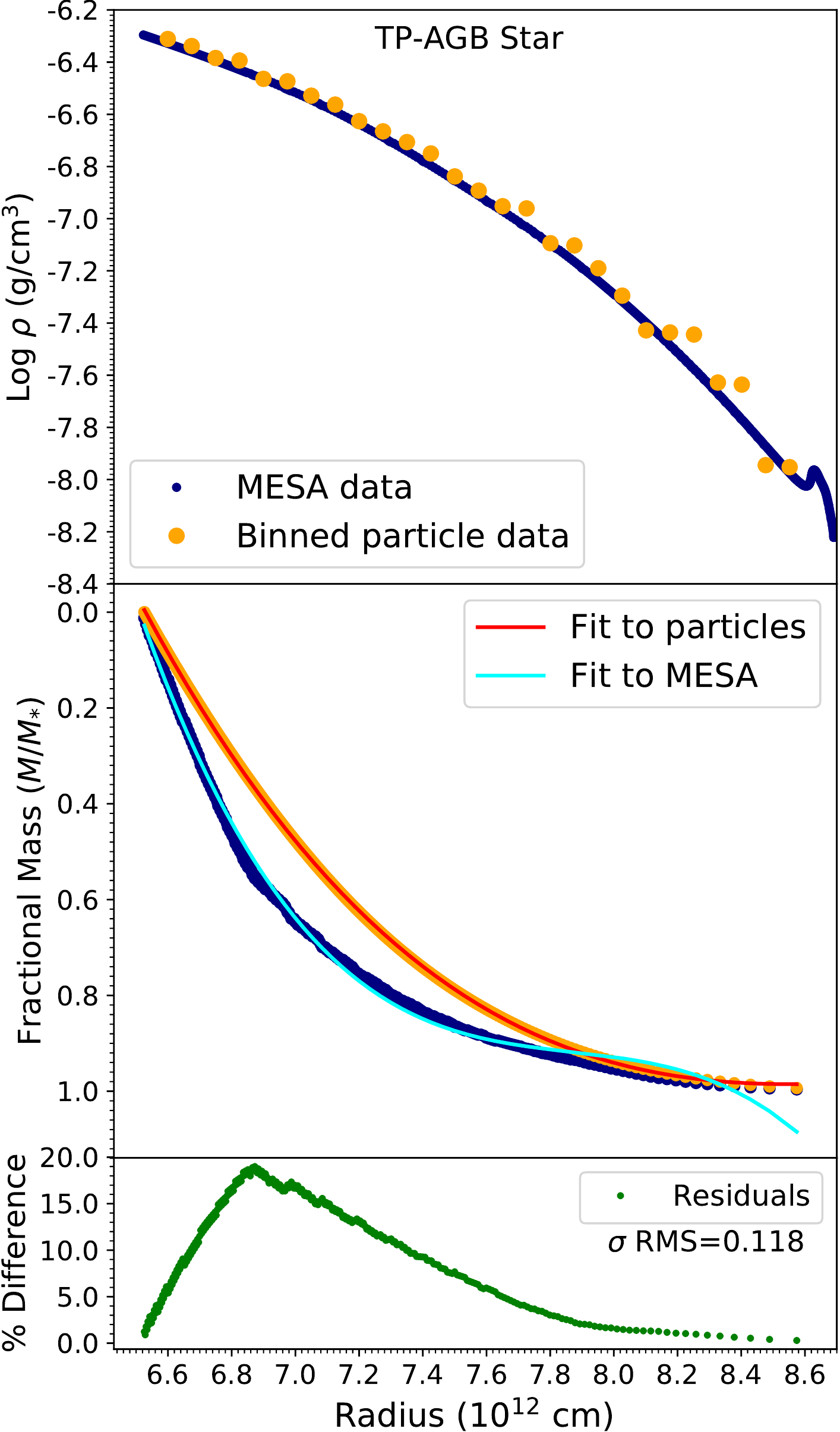}
\caption[Recovery statistics for TP-AGB model]{Same as Figure \ref{starfig1a}, but for a TP-AGB model.}
\label{starfig2b}
\end{figure}

\begin{figure}
\centering
\includegraphics[width=\linewidth]{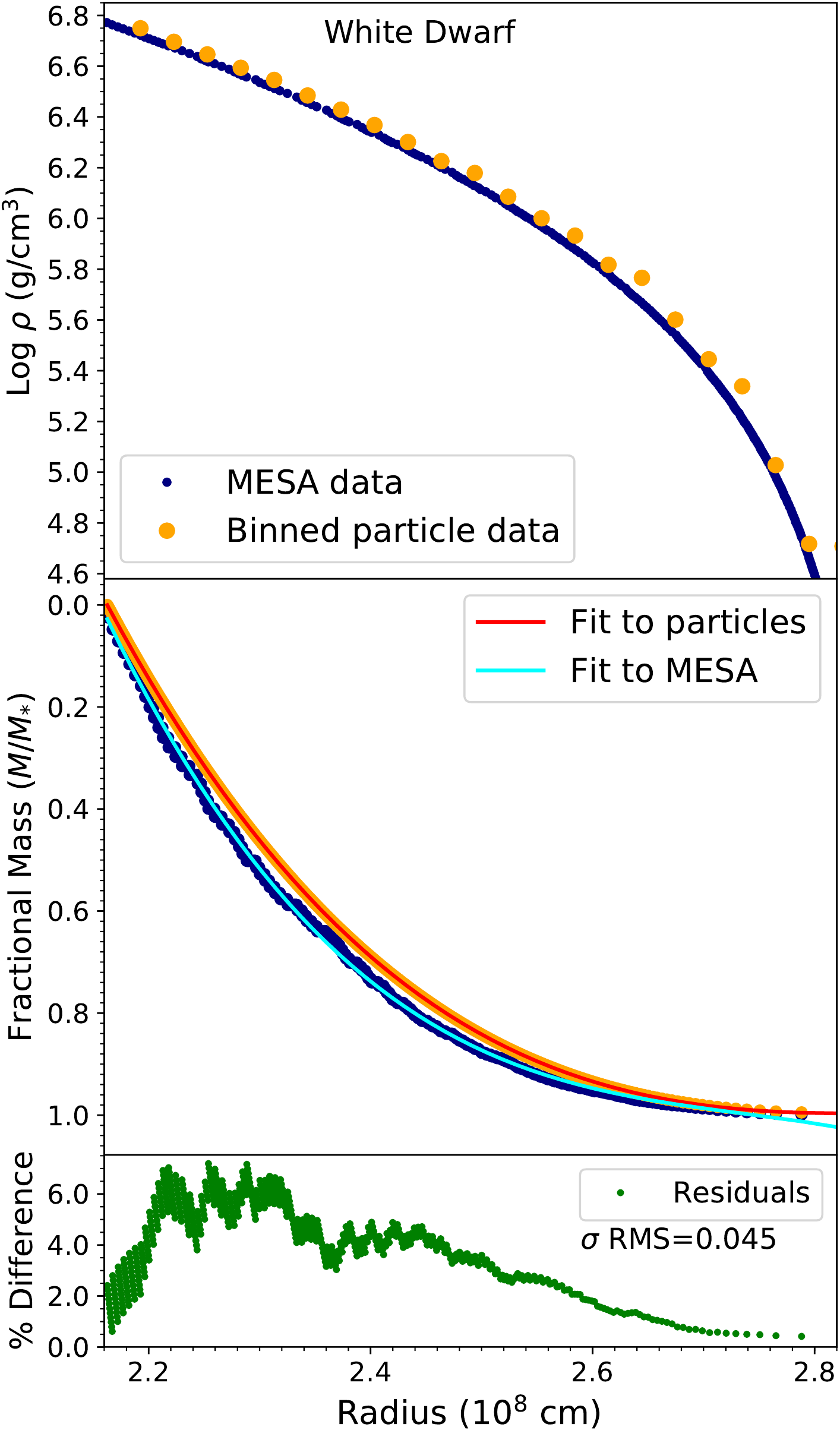}
\caption[Recovery statistics for white dwarf model]{Same as Figure \ref{starfig1a}, but for a white dwarf model.}
\label{starfig3}
\end{figure}


\subsection{Test Cases}
\begin{unenumerate}
\def\item[]#1{\smallbreak\noindent{#1}\enspace}
\item[] {Model 1: Solar-like Main Sequence Star}
This is a standard model with solar mass ($1.0\, M_{\odot}$) and metallicity ($Z=0.02$) which {terminates $5$ Gyr} after collapse onto the {zero-age} main sequence. 
{The age, initial and final masses, metal abundance, radius, luminosity, and temperature are given for this and subsequent models in Table \ref{stars}. }
The profile roughly reflects the interior structure of the Sun. Solar-mass stars transport energy via convection in their outer layers and via radiation in their cores. {Though the Sun at its current age is not especially dynamic, interactions with the convective envelope produce mixing events in the later stages of a star's life, which can lead to instability. This, and the fact that the Sun is used as a calibrator for 1-D stellar models, make this a particularly important test case \citep{Joyce2015, Joyce2018a, Joyce2018b}. }

\item[] {Model 2: Solar-like Red Giant}
This model has a mass of $1\, M_{\odot}$ and solar metallicity. It terminates {at 12 Gyr,} after it has left the main sequence and ascended onto the red giant branch. Red giants of roughly this mass and age are common in stellar populations such as globular clusters, and they frequently appear in multi-star systems. This model is a useful template for the canonical red giant in a binary system.

\item[] {Model 3: Supergiant}
This model has an initial mass of $90\,M_{\odot}$, solar metallicity, and terminates in the middle of its red giant phase. Because of its mass, the interior physics and sub-surface density structure differ considerably from Model 2. At its termination, the interior will have a similar configuration in terms of convective versus radiative zones, but this model contains nested hydrogen, helium, carbon, and oxygen burning regions. Most importantly, the outer layers of this star are highly radially extended. While the outermost 5\% of mass is contained within the outermost 20\% of radius for a solar-like red giant, the same radial proportion holds less than 0.5\% of the mass in this case.
This model {is guided by the inlist used to generate Figure 44 in \citet{MESAIII}, and  our parameter choices are informed by those of the ``90M\_logT\_9.35.mod'' model in the high-mass test suite of MESA version 8118. We include the inlist adapted for mesa-r10398} with \m2h.

\item[] {Model 4: Thermally Pulsating AGB Star}
This model has an initial mass of $2.55\,M_{\odot}$ and metallicity $Z=0.01$. 
It terminates after the third dredge-up during the late thermal pulse phase along the asymptotic giant branch. At this point in its evolution, the star's pulsations {have caused a small amount of mass loss, and so the initial mass is slightly} higher than the mass integrated in its density profile. {This model adopts the opacities based on \citet{Asplund09} and the ``Mesa\_49.net'' nuclear reaction grid.}

The {metallicity, opacity prescription, and nuclear network} deviate from what is used in other models because {this inlist} is based {on an experimental inlist designed specifically to achieve good convergence on post-third dredge-up (3DU) thermal pulses; this is a
notoriously difficult evolutionary phase to model correctly (see, for example, \citealt{TUMi}). It also borrows overshoot and mixing prescriptions from inlists for TP-AGB models published by \citet{Tashibu}, and invokes other suggestions from the MESA collaboration directly (see the inlist and \citealt{MESAIII} for more detail).}

\item[] {Model 5: White Dwarf}
This is a solar-mass white dwarf, {and it is the only one in our suite that uses a pre-built MESA model. Specifically, we use  
\verb|wd3_1Msolar.mod|, one of several white dwarf models included with MESA 10398. 
The parameter specifications required to produce a reasonable profile for the outer layers of a white dwarf are very different from those of the previous models due to the fundamental difference between white dwarfs and the outer regions of ``living'' stars, namely, the equation of state.
A non-differentiable region may emerge due to the disruption in the stellar profile where the equations of state governing different regions, e.g., cores versus atmospheres, meet, or it may be the consequence of perfectly physical discontinuities in the chemical composition of the star.} 
For a star undergoing active nuclear fusion, such discontinuities emerge deep in the interior and would not affect our modeling of the outer layers. For a white dwarf, however, discontinuities occur very near the surface, as white dwarfs have inert cores under incredibly thin atmospheres. 
{This model includes Type2 opacities and invokes the ``co\_burn.net'' nuclear reaction network, which accounts for additional Carbon/Oxygen burning. This model also includes an accretion rate of $10^{-9}$ M$_{\odot}$/yr during the white dwarf phase, which accounts for the small increase in mass over its initial, solar value.}
A (non-MESA) stellar structure model provided by Maurizio Salaris \citep{BASTI06,SalarisWD} was used to guide additional parameter selections for this model.  
\end{unenumerate}

Table \ref{stars} summarizes basic physical attributes of the models, including their stopping criteria.
This set is by no means exhaustive, but does cover key regions of the HR diagram and includes types of stars that are commonly modeled in hydrodynamics simulations.

\begin{table*} 
\centering 
\small
\caption{Summary of Physical Properties of Modeled Stars}

\begin{tabular}{l ll  rrr rr  lc } 
\hline\hline
Type &

$M_{\text{i}}$ &
$Z_{\text{i}}$ &

Age &
$M_\text{f}$ &
$R$ &
$\log L$&
$T_\text{eff}$&

Stop Condition &
Pre-built model?
\\ \hline
\vbox to 10pt{}Main Sequence &  1.0 	& 0.02 & 5.0 & 1.0 & 1.0 & 0.0 & 5760	& max\_age = $3.0$d9 		& no \\
 Red Giant 	  				& 	1.0		& 0.02 & 12.0 & 1.0 & 3.44 & 0.76 & 4820	& max\_age = $12.0$d9 		& no \\
 Red Supergiant  			&  	{90}& 0.02 & 0.00015 & 90 & 9.29 & 6.14 &65100	& H-burning limit 			& {no} \\
 TP-AGB  					&   2.55 	& 0.02 & 0.77 & 2.54 & 125& 3.28 & 3400 & stop\_at\_TP=True 		& no \\
 White Dwarf 		 	 	&   1.0		& 0.02 & $>15$& 1.3 & 0.004& -0.48 &	68400& Nuclear burning limit 	& yes \\
\hline
\end{tabular}
\tablecomments{Basic physical attributes of the test models. Masses are in M$_{\odot}$. 
{ $Z$ is the dimensionless mass fraction of metals. Ages are in Gyr.
Radii are in units of R$_{\odot}$. Log $L$ are in units of $\log$ L/L$_{\odot}$. Effective temperatures are in K.} Stopping criteria are used to halt the evolution of the model at the correct phase. The first two models are stopped at certain ages corresponding to the correct evolutionary phase. Models 3 and 5 are stopped according to limits on their nuclear reactions, which indicate the correct evolutionary phases. Model~4 uses a flag specifically for stopping evolution during a thermal pulse.}
\label{stars}
\end{table*}


\begin{figure}[h!] 
\centering
\includegraphics[width=\linewidth]{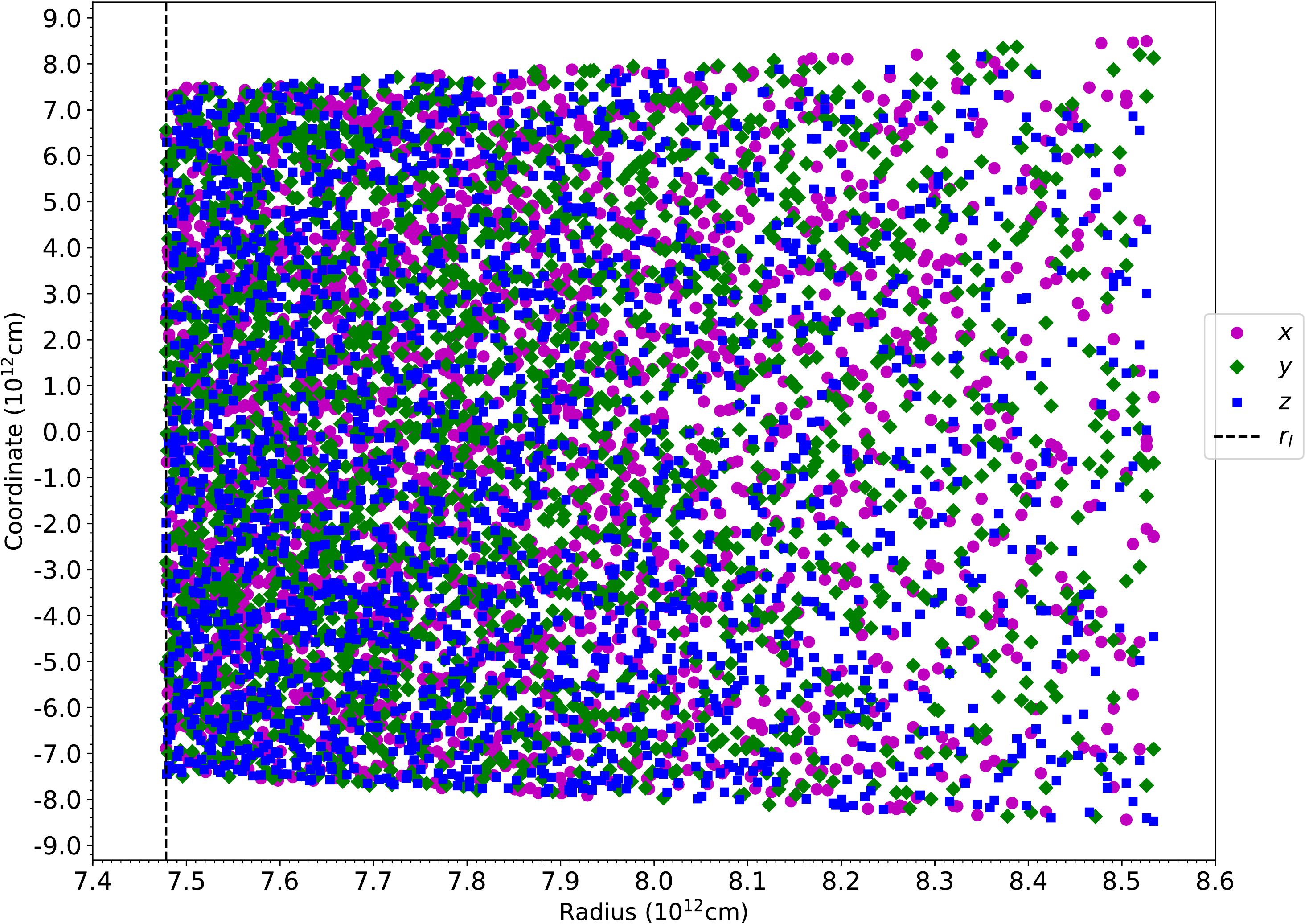}
\caption[Cross section of an \m2h-generated particle distribution]{%
A sample (every $200^{\text{th}}$ particle of the array of ${\sim}460,000$) of the $x$ (pink), $y$ (green), and $z$ (blue) components from a particle distribution representing the TP-AGB model are shown against radial coordinate $r=\sqrt{x^2 + y^2 + z^2}$. The vertical dashed line indicates the inner-most radius that was captured by the TP-AGB model's depth cut of 1\%. This slice corresponds to Figure \ref{starfig2b}.
}
\label{crosssec}
\end{figure}

\begin{figure} 
\centering
\includegraphics[width=0.78\linewidth]{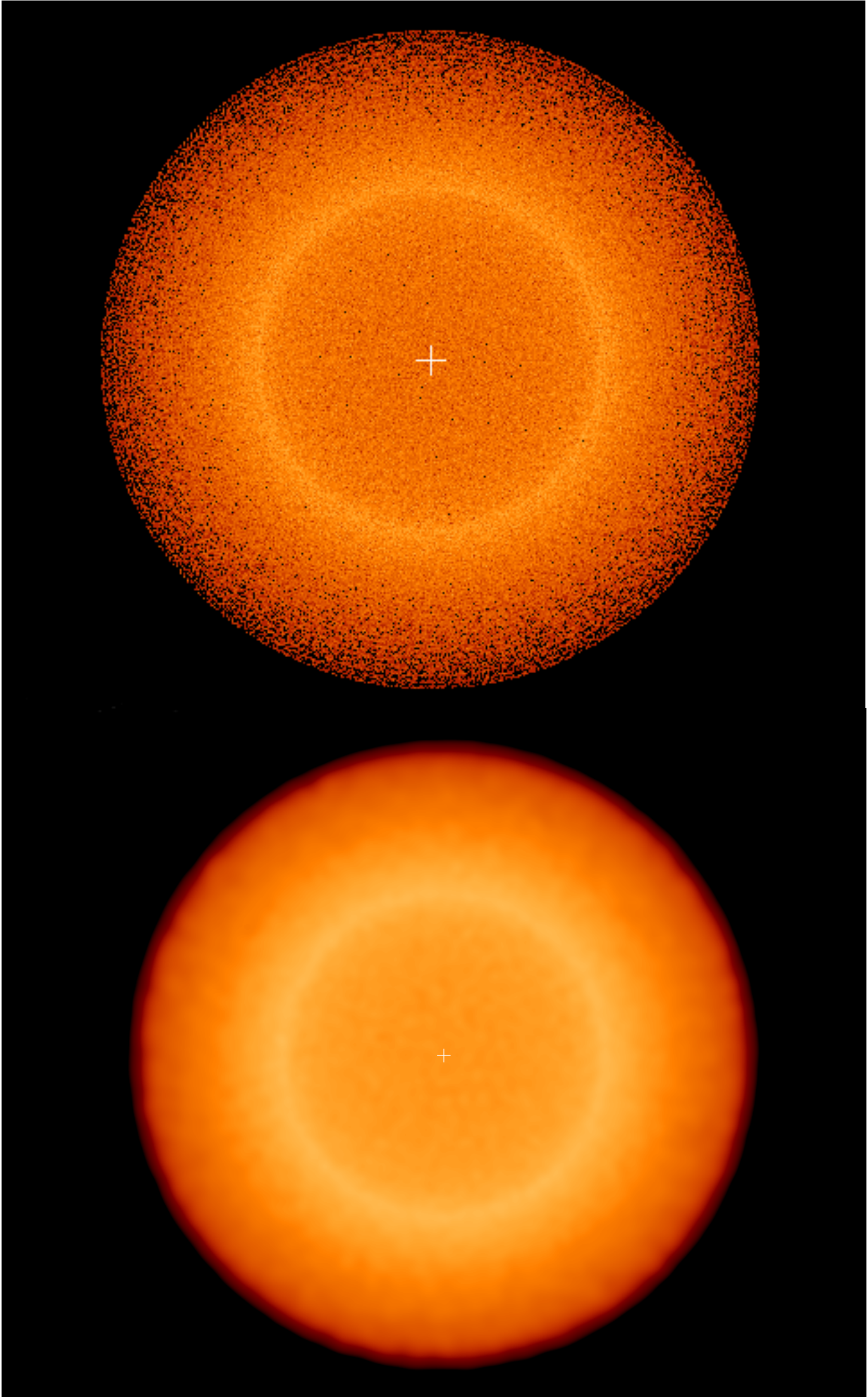}
\caption[3-D map of output particle distribution]{\footnotesize
Top: A \m2h-generated particle distribution, as rendered by the SPH viewer {gadgetviewer}, for the TP-AGB model.
The region modeled extends from the central, yellow ring to the exterior. The center is hollow (as demonstrated in Figure \ref{crosssec}); particles which appear in the center are physically located at $r \ge \sim7.5 \times 10^{12}$ cm, but are displayed in the plane of the page. 
Bottom: This panel shows the same, but a smoothed-density projection has been applied. Color coding reflects density, where yellow regions are most dense and red regions are least dense. 
Note that in the case of the smoothed distribution, the density coloring reflects the presence of the large point mass we have placed at the center of the star.}
\label{final_dist}
\end{figure}

%
%
\subsection{{Validation: MESA to Particle Comparison} }

We quantify the consistency between the MESA-generated, source density profiles and the density profiles estimated from particle distributions recovered from \m2h. Figures \ref{starfig1a}--\ref{starfig3} contain the following:
\vskip1.7ex

\noindent (1) In the upper panels, semi-log density profiles of the MESA-generated source data and particle data are shown in physical units. A density distribution is estimated from the particle data {using bin numbers ranging from $20$ to $30$ among the test cases.}

\vskip1.7ex
\noindent (2) In the central panels, normalized cumulative mass distributions for source and particle data are shown. 
They are inverted to reflect the curvature of the (unlogged) density profiles and normalized to remove dependence on particle mass. We use $m_p=10^{-7}$ in all cases {but the supergiant, which uses $m_p=10^{-6}$}, but this is an arbitrary choice and does not impact the shape of the density and mass distributions. 
A third-degree polynomial 
\begin{equation}
\rho_f(r)=Ar^3 + Br^2 + Cr + D
\label{poly}
\end{equation}
is applied to each curve to allow for comparison between $\rho(r)$ predicted by each distribution.

\vskip1.7ex
\noindent (3) The lower panels present residuals in terms of percent error,\ $100\times|\rho_f(r)-\rho_g(r)|$, as a function of radius,
{where $\rho_f(r)$ is a fit of the form \eqref{poly} to the source density profile and $\rho_g(r)$ is the binned particle data.}

\vskip1.7ex
\noindent (4) The error is also given as a single RMS score in the lower panels, 
$$\sigma_{\text{rms}} =\sqrt{ \Sigma^N_k \frac{ ( \rho_f(r) -\rho_g(r) )^2 }{N_{\text{shells}}} },$$
calculated only over the radial extent shown.

\begin{table*} 
\centering 
\caption{\m2h Run Time Parameters and Goodness of Recovery for Test Models}
{\small 
\begin{tabular}{l cccc l cccc}  
\hline\hline
Star  &
$N$   &
$m_p$ &
$R_{\star}$ &
$M_{\star}$ &
$\Delta_{r,\text{initial}}$  & 
$N_{\text{shells}}$ &
$n_p$ &\
$t_{\text{gen}}$ & 
$\sigma_{\text{rms}}$ 
\\ \hline
Standard Solar 	& 8  &  $1\times10^{-7}$  &  0.75  &  0.0167  &  1.00$\times 10^{8}$  &  441  &  338,688    &  1.311  &  0.018  \\ 
Solar Red Giant & 8  &  $1\times10^{-7}$  &  0.75  &  0.0650  &  1.00$\times 10^{8}$  &  852  &  654,336    &  4.039  &  0.022  \\ 
Red Supergiant 	& 8  &  $1\times10^{-6}$  &  0.75  &  0.0034  &  1.00$\times 10^{11}$ &  453  &  347,904  	&  6.671  &  0.020  \\ 
TP-AGB 			& 8  &  $5\times10^{-7}$  &  0.75  &  0.0449  &  1.00$\times 10^{10}$ &  298  &  228,864    &  6.652  &  0.118  \\ 
White Dwarf 	& 8  &  $1\times10^{-7}$  &  0.75  &  0.0330  &  1.00$\times 10^{5}$  &  556  &  427,008    &  5.128  &  0.045  \\ 
\hline
\hline
\end{tabular}
}
\tablecomments{
Computational and physical features of each test model are shown for $r_{\text{depth}}=0.75R_{\star}$ and $\delta_\mathrm{TOL} = 0.01$, corresponding to Figures \ref{starfig1a} through \ref{starfig3}. Particle masses are in units of $M_{\odot}$. $N$ is the HEALPix integer. Initial step sizes are in physical units (cm). $t_{\text{gen}}$ is the generation time for the NR coordinates, in hours. Generation times for IC files are negligible, typically on the order of 5 to 10 seconds.}
\label{recovery} 
\end{table*}

{We generate sets of models mapping the outer $25$\%, $50$\% and entire star by radius ($r_{\text{depth}}=0.75, 0.5$ and $0.0$ in the convention of the configuration files). Figures \ref{starfig1a} through \ref{starfig3} show results for $r_{\text{depth}}$ fixed to $0.75R_{\star}$ and $\delta_\mathrm{TOL} = 0.01$. We fix the HEALPix integer to $N=8$ in each case {but allow $m_p$ to vary among models}. 
These demonstrate the goodness of recovery for models of varying structure, spanning $18$ dex in density, constrained by computation times on the order of hours and tractable particle numbers ($<0.5$ million).}
Table \ref{recovery} shows the numbers of shells and particles needed to generate Figures \ref{starfig1a}--\ref{starfig3}, the radial and mass percentages modeled in each case, the computation times, and {the RMS measure of} goodness of fit.

{The maximal discrepancy between input and output density profiles with these parameter settings is less than roughly~10\% for all models, and less than 5\% for all but the TP-AGB star. The RMS error for the TP-AGB star is more than double the next-most discrepant model, the white dwarf, whose RMS error is itself more than double that of the other three test models. One factor contributing to these differences in agreement is the comparatively larger range of densities encompassed in the outer 25\% of the AGB star and white dwarf---about $2.2$ dex versus $1.4$ dex for the main sequence star, red giant, and supergiant. 

As shown in Figure \ref{starfig2b}, \m2h struggles most with resolving the outermost portion of the stellar atmosphere, which is unsurprising given the sparseness of this region. If one's primary goal is to preserve the structure of the outer layers with high fidelity, a smaller particle mass and shallower $r_{\text{depth}}$ would be more appropriate choices than those used in Table \ref{recovery}.

We note that some subtleties, such as the hook in the AGB star's outer-most density profile in the upper panel of Figure \ref{starfig2b}, may be lost at the expense of large choices for $m_p$ / lower particle numbers / faster integration times.  }
Greater consistency between the input and \m2h-rendered density profiles is always achievable with larger particle numbers and longer computation times, but distributions with extremely large particle counts will be intractable for the user as well as for most hydrodynamics codes. 
There is also a point of diminishing returns when balancing computation time against reduction in $\sigma_{\text{RMS}}$, and it is up to the user to decide which parameters are appropriate for her problem. Those presented here serve only as a guide.

In addition to the tests summarized in Table \ref{recovery}, we generate distributions over the same radial span with equivalent parameter values using a weaker tolerance, $\delta_\mathrm{TOL} = 0.05$, and note the difference in computation time. For the main sequence model, the execution takes $0.16$ hr at $\delta_\mathrm{TOL} = 0.05$ versus $0.74$ hr at $\delta_\mathrm{TOL} = 0.01$. For the red giant, the run time drops to $0.99$ hr; for the supergiant, AGB star, and white dwarf, the run times drop to $1.31$, $0.73$, and $1.22$ hr, respectively, from the times given in Table \ref{recovery}.
The corresponding loss in fidelity for such shallow radial penetration depths is high, but increasing $\delta_\mathrm{TOL}$ when integrating half or all of the radial profile has a less significant impact on the overall shape of the recovered density profile. 
The diminished resolution will impact low-density regions most. In such a case where one is interested in the density structure of the entire star rather than strictly the outer layers, this loss may be worth the decrease in computation time.

We can understand the impact of the choice in particle mass on computation time by comparing the run time values in Table \ref{recovery} to those of the modified models used to generate the test ICs for verification with Phantom (these tests are discussed in more detail in Section \ref{section:Phantom}). 
Using $\delta_\mathrm{TOL}=0.01$ and particle masses of $10^{-5} M_{\odot}$ for the main sequence star, red giant, and white dwarf, and $10^{-4} M_{\odot}$ for the AGB star and supergiant, \m2h's search for radial solutions to equation \eqref{int} over the majority of the stellar profile ($r_{\text{depth}} \rightarrow 0$) takes $\le2$ hours each model. Calculation times for the red supergiant and white dwarf are longest, whereas the NR coordinates for the main sequence star, red giant, and AGB star take less than 40 minutes to generate. 

We find that, so long as the integration tolerance is maintained, choosing higher values of $m_p$ generously improves the speed of the NR routine without heavily impacting the shape of the damped density profile at the end of a simulation with Phantom. 
We find that precision on the location of the shell radii is more important for preserving agreement with the initial MESA profile than resolution in particle mass, so it is preferable to lower $\delta_\mathrm{TOL}$ than to lower $m_p$ when seeking improvements in fidelity without huge increases in computation time. 
We note, however, that these are appropriate trade-offs only for distributions that encompass most of the star, by radius.}

\subsection{{Validation: 3-D Verification} }
\label{section:Phantom}
A cross-section of an \m2h-generated particle distribution for the TP-AGB model, as an example, is shown in Figure \ref{crosssec}, and the 3-D distribution is shown in Figure \ref{final_dist}. The lower panel of Figure \ref{final_dist} invokes a smoothing kernel. Distributions for all test models look similar when viewed with an SPH-compatible program \citep{gadgetviewer, SPLASH}.

{We further analyze each of the five test distributions using the Phantom smoothed-particle hydrodynamics code \citep{Phantom}. 
It is easiest to achieve stable distributions in Phantom as $r_{\text{depth}}$ approaches zero, but the resulting particle numbers for the same combinations of $m_p$ and $\delta_\mathrm{TOL}$ given in Table \ref{recovery} exceed $10$ million for some models (e.g., the white dwarf). While it is possible to calculate the evolution of 10 million particles with Phantom, the computing times extend into weeks. 
As rigorous hydro\-dynamical analysis is beyond the scope of this paper, we perform the 3-D tests using lower resolution versions of the models presented in Table \ref{recovery}.
With the exception of larger values of $m_p$ and penetration depths approaching zero, all parameters of the ICs run with Phantom are equivalent.

We evolve the distributions for 10 dynamical timescales ($\tau_{\text{dyn}}$), following the damping prescription of \citet{Ohlmann}'s equation (9), implemented in Phantom by \citet{ReichardtThesis}:
\begin{equation}
  \tau(t) = \begin{cases}
    \tau_1, \quad & t < 2t_\mathrm{dyn}, \\
    \tau_1 \left( \frac{\tau_2}{\tau_1} 
    \right)^{\frac{t-2t_\mathrm{dyn}}{3t_\mathrm{dyn}}}, \quad & 2t_\mathrm{dyn}
    < t < 5t_\mathrm{dyn}, \\
    \infty, \quad & t > 5t_\mathrm{dyn}. 
  \end{cases}
  \label{eq:relaxationtime}
\end{equation}
Dynamical timescales are proportional to density and thus vary among the models. Very rough estimates are given by
\begin{equation*}
t_\mathrm{freefall} \approx \frac{2100}{\sqrt{\rho}},
\end{equation*}
where the numerator is in seconds and the denominator is in g/cm$^3$. The order of $t_\mathrm{dyn}$ for each model is estimated from $\rho_{\star} \sim \frac{M_{\star}}{R_{\star}^3}$ and given in Table \ref{3Dpar}.

We consider the initial conditions to be valid as long as a few conditions are satisfied. First, all physical attributes assigned via \m2h must be read and correctly understood by Phantom. Second, the shape of the density profile must be preserved at $t=10 t_\mathrm{dyn}$.
Last, if the velocity dispersion is small after evolution without damping for $5 t_\mathrm{dyn}$, we consider the final configuration to be in hydrostatic equilibrium. However, we do not make this a strict criterion for claiming consistency, as there may be valid physical reasons for instability after several dynamical timescales for some of our models. }

\begin{figure}[h!] 
\centering
\includegraphics[width=\linewidth]{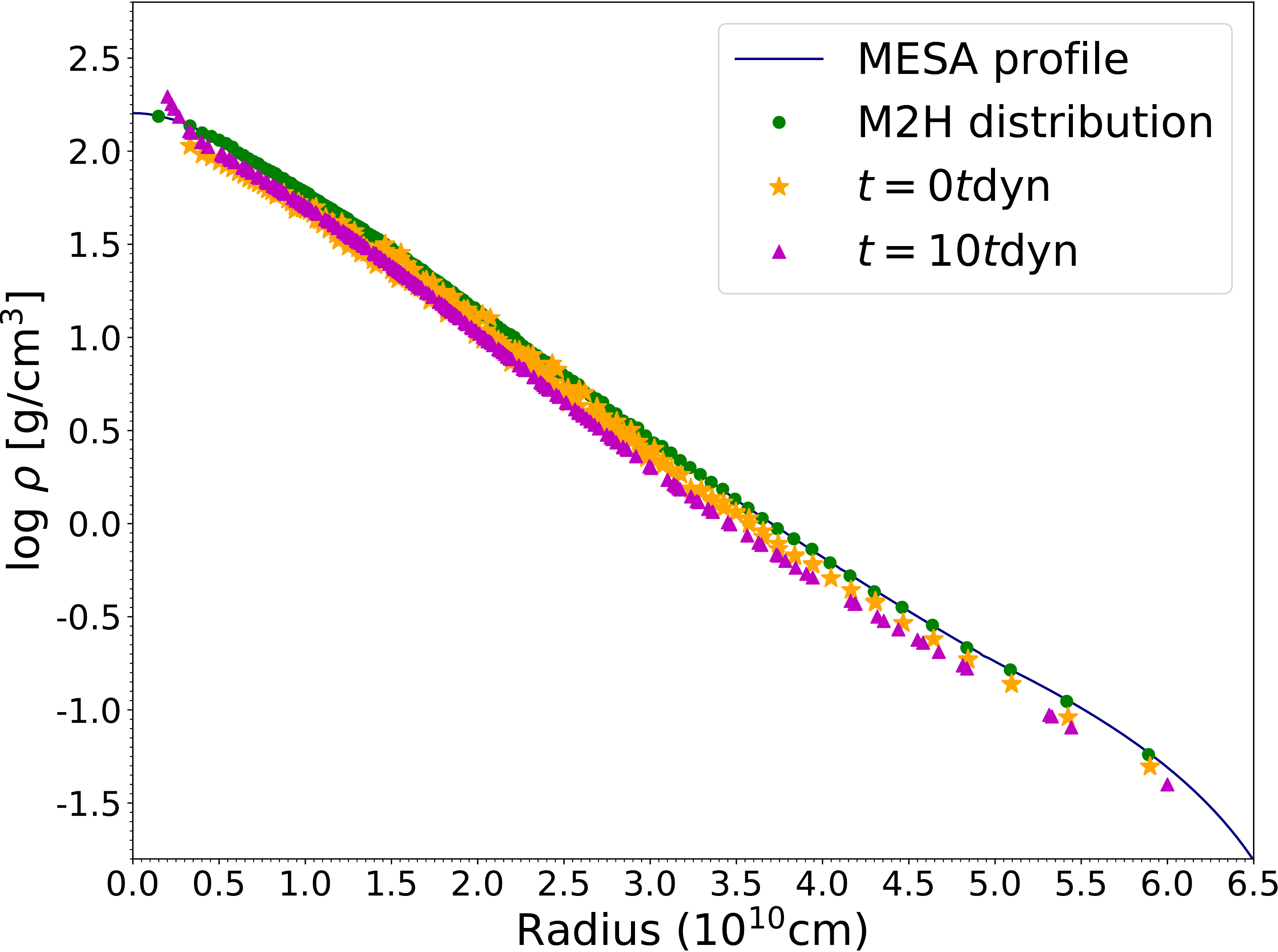}
\caption[comparison in 1D]{%
{For the solar-like, main sequence star: (1) the MESA profile is shown as a blue line; (2) the output from \m2h reduced to 1-D is shown in green dots; (3) the distribution reloaded from Phantom at $t = 0 t_{\text{dyn}}$ is shown in yellow stars; (4) the stable particle distribution reloaded from Phantom at $t= 10 t_{\mathrm{dyn}}$ is shown in purple triangles.}
}
\label{1Dcompare}
\end{figure}

\begin{figure}[h!] 
\centering
\includegraphics[width=\linewidth]{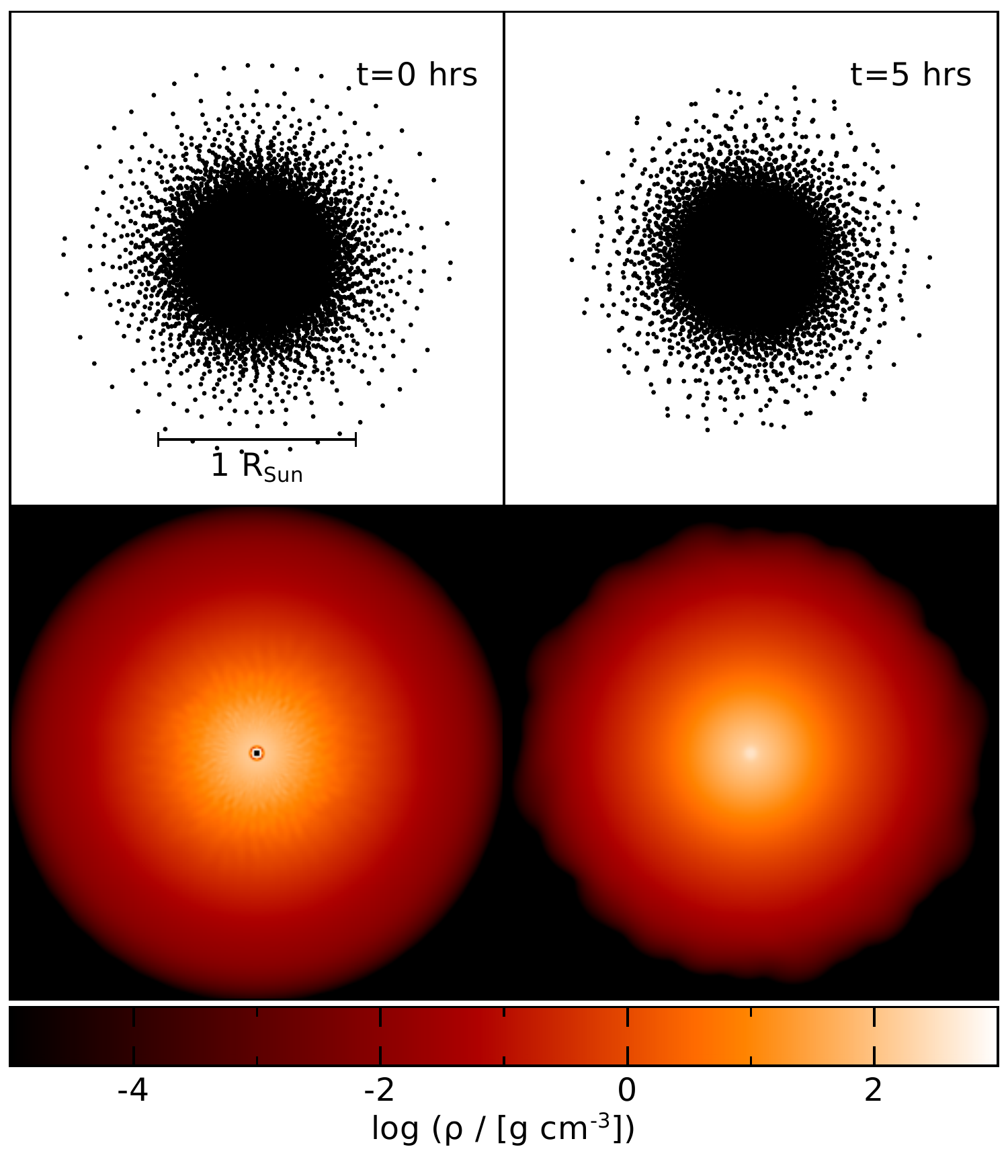}
\caption[]{%
For the main sequence model, the left panels show features of the particle distribution at $t = 0 t_{\text{dyn}}$, and the right panels at $t = 10 t_{\text{dyn}}$,  in hours.
Top: Physical particle distribution in the cross section.
Bottom: Density field in the $XY$ plane. 
}
\label{SPHsun}
\end{figure}

\begin{figure}[h!] 
\centering
\includegraphics[width=\linewidth]{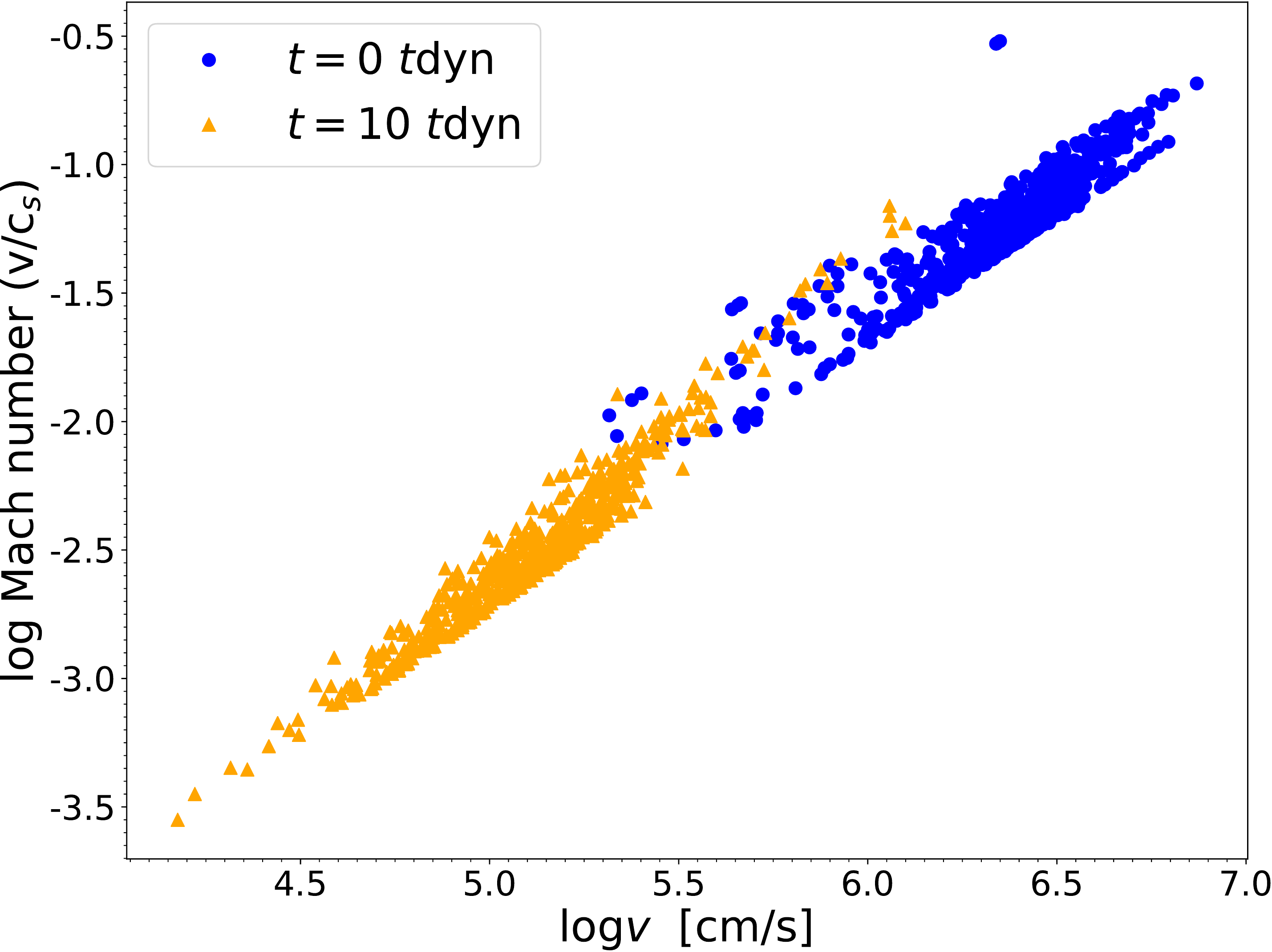}
\caption[]{%
Mach number $v/c_s$ versus velocity is shown for a subsample of the particles comprising the 3-D main sequence model. A snapshot at $t=0t_{\text{dyn}}$ is shown in blue circles; at $t=10t_{\text{dyn}}$, in orange triangles.}
\label{velocity}
\end{figure}

The relationship between (1) the initial MESA profile, (2) the 1-D approximation to the particle distribution generated with \m2h, (3) the \m2h output reloaded from Phantom ($t_{\text{dyn}}=0$), and (4) the converged particle distribution at $10 t_{\mathrm{dyn}}$ for the main sequence test model is shown in Figure \ref{1Dcompare}. By visual estimation, there is excellent agreement among all four density profiles in 1-D.

We note some flattening of the profile occurs between Phantom's interpretation of the \m2h particle distribution and between 0 and 10 $t_{\text{dyn}}$, and that the inner and outer boundaries of the star are less well resolved than the central region. The difference between the raw output from \m2h and the distribution reloaded from Phantom at $t=0$ is that, in the former case, the output from \m2h is converted back to radial coordinates and assigned the nearest density from the MESA profile directly. In the latter, the density values assigned to each particle reflect Phantom's own calculations based on nearest neighbors, using a smoothing kernel. 

Appropriate smoothing lengths for the gas particles are computed in \m2h according to 
\begin{equation*}
h_i=1.2 \times (m_p/\rho_{i})^{1/3},
\end{equation*}
where $i$ is the index in MESA of the radius corresponding to the particle's shell radius, and $\rho_i$ is the corresponding MESA density value.

As one can see in Figure \ref{1Dcompare}, there is a spread among the yellow particles, reflecting variations in density estimates for particles situated on the same shell. While these fluctuations quickly dissipate during the damping procedure, they can also be mitigated initially by increasing the precision on equality \eqref{int} (i.e., lowering $\delta_{\text{TOL}}$ and/or $m_p$) when generating the NR file.

Figure \ref{SPHsun} shows $XY$ cross sections of the 3-D particle distributions corresponding to the $t=0 t_\mathrm{dyn}$ and $t=10 t_\mathrm{dyn}$ curves in Figure \ref{1Dcompare}. Here, we see that the distribution at  $10 t_\mathrm{dyn}$ has settled into a configuration that appears more random and slightly less radially extended. In reality, some flattening of the density profile and radial extension occurs for all models in the 3-D simulations.

Figure \ref{SPHsun} also shows a small black region in the center of the density field in the $t=0$ distribution---suggesting a central density of zero---which is subsequently replaced by a value close to the true stellar central density at $t=10$. 
In Phantom, the core mass is modeled as a sink particle which interacts only gravitationally with the gas particles, and the softening length ($h_{\mathrm{sink}}$) of the sink particle is calibrated to be half of the the inner-most radius of the gas shells. 
This is placed at the physical center of the distribution. 
The ``filling in'' of the density field at the center over simulation time is caused by Phantom's attempt to smoothly connect the fields of the gas particles and the central gravity well (for details on the mathematics of how this is achieved, see \citealt{Phantom}).

Even in cases where nearly all of the stellar radius is captured by \m2h, the sink particle will exert a force on the gas particles proportional to the integrated density in the center of the star. As such, the sink particle is crucial in the balancing of forces required to achieve hydrostatic equilibrium, and the less work Phantom needs to put into smoothly connecting the pressure of gas particle distribution with the gravitational exertion of the sink particle, the more stable its solution will be. This is why it is helpful, in terms of computation time and accuracy on the 3-D end, to set $r_{\text{depth}}$ close to zero in \m2h.
The trade-off will be either lower fidelity in the outer layers of the star or large particle numbers; prioritizing these attributes is, once again, a matter of preference, and best options will vary depending on the problem.

To further demonstrate that particle configurations generated with \m2h behave reasonably in hydrodynamic simulations, we examine the evolution of the velocity distribution in the main sequence star, a system we know should not exhibit any mass loss, radial expansion, or dynamical behavior.
Figure \ref{velocity} shows Mach number (velocity over local sound speed, $c_s$) versus velocity for a subset of the particles in the main sequence model at the beginning and end of its evolution with Phantom. This clearly demonstrates the trend towards smaller particle velocities, on average, with increasing evolutionary time---inclusive of evolution over $5$ dynamical timescales without damping.
The RMS velocities for $t=0$ and $t=10$ are $2.8\times10^6$ cm/s and $2.1\times10^5$ cm/s, respectively.
Table \ref{3Dpar} summarizes the 3-D simulation parameters for each test model.

In most cases, the number of particles that become gravitationally unbound from the star (on our simulation timescale) is zero or negligible; however, the AGB model in particular is prone to mass dissipation and rapid radial expansion. This is not necessarily unphysical, as the MESA profile for this model does characterize a star in the midst of dynamical activity, and the internal energy associated with this state is propagated through \m2h into the hydrodynamic ICs. 
The generation of hydrodynamically stable models of AGB stars and other radially extended giants that do not require external damping or artificial forces remains an active area of research, and this is not a problem we attempt to solve in this study. 
However, though rigorous assessments of the evolution of these systems are beyond the scope of this paper, we believe \m2h will be a valuable tool in the search for solutions to these and other outstanding problems in stellar hydrodynamics.

As a baseline assessment, we check that the shapes of the \m2h-generated distributions remain intact over $10 t_{\mathrm{dyn}}$ and compare the \m2h distributions to polytropic test models (available by using the ``star'' Makefile option in Phantom; see \citealt{Phantom}) which use the same number of particles, maximum radius, EOS/adiabatic index, and central density. In all cases, the \m2h distributions and polytropic distributions show similar evolution over $10 t_\mathrm{dyn}$. Both undergo some radial expansion, typically on the order of $5\text{--}20\% R_{\star}$. However, unlike the polytropic models, the \m2h distributions are derived from MESA density profiles directly. 

We do note that in the case of the white dwarf, there is little difference in the evolution of a \m2h distribution versus a polytrope generated with Phantom, and it is more time-consuming to use \m2h. It is also perhaps most appropriate to use the ``whitedwarf'' Makefile setup in Phantom, which involves more sophisticated nuclear reaction networks that are not yet publicly available. Our tool nonetheless renders white dwarf density profiles effectively, though other means may be more appropriate for this type of star.

\begin{table*} 
\centering 
\caption{Parameters for \m2h Distributions Tested with Phantom}
{\small 
\begin{tabular}{ lc lc  lll l } 
\hline\hline
Star &

$m_p$&
$N_p$ &

$M_{\star}$ (g) &
$R_{\star}$ (cm) &

$\rho_{\mathrm{avg}}$ (g/cm$^3$) &
$t_{\mathrm{dyn}}$ (s) &
Phantom EOS 
\\ \hline
Standard Solar 	& $10^{-5}$ & 102,144  & $1.98\times10^{33}$  	&  $7.02\times10^{10}$  &  $1.37\times10^{0}$  &  $1.80\times 10^{3}$  & Adiabatic; $\gamma=5/3$  \\ 
Solar Red Giant & $10^{-5}$ & 102,144 	& $1.98\times10^{33}$ 	&  $2.40\times10^{11}$  &  $3.46\times10^{-2}$ &  $1.13\times 10^{4}$  & Adiabatic; $\gamma=5/3$ \\ 
Red Supergiant 	& $10^{-4}$ & 64,512	& $1.78\times10^{35}$ 	&  $1.18\times10^{14}$  &  $2.57\times10^{-8}$ &  $1.31\times 10^{7}$  & Adiabatic; $\gamma=5/3$ \\ 
TP-AGB 			& $10^{-5}$ & 193,536 	& $5.05\times10^{33}$	&  $8.69\times10^{12}$  &  $1.84\times10^{-6}$ &  $1.55\times 10^{6}$  & Adiabatic; $\gamma=5/3$ \\ 
White Dwarf 	& $10^{-5}$ & 134,400  & $1.98\times10^{33}$ 	&  $2.78\times10^{8}$   &  $2.20\times10^{7}$  &  $4.48\times 10^{-1}$ & Helmholtz \\ 
 \hline
\end{tabular}
} 
\tablecomments{%
In all cases, $\delta_{\text{TOL}}=0.01$. $r_{\text{depth}}$ is approximately zero for all models, with the exception of the supergiant, which uses $r_{\text{depth}}=0.35$ due to the difficulty of resolving the extreme densities in the stellar core using a particle mass that is suitable for the rest of the layers. When $r_{\text{depth}}=0.0$ is used for the supergiant, the \m2h solution time jumps from approximately 1.6 hr to 19 hr, and the particle number increases from $\approx65000$ to more than 1 million. 
}
\label{3Dpar} 
\end{table*}

\section{Summary and Conclusions}
We have presented \m2h, a tool for generating 
initial conditions for hydrodynamical simulations through the translation of 1-D, SSEC-based density profiles to 3-D particle distributions.
We have presented a technique for generating $N,R$ coordinates, modeled in part after similar efforts by \citet{Pakmor2012} and \citet{Ohlmann}. We have explained \m2h's operational features, primary algorithm, and underlying numerical methods.

We have verified \m2h's functionality on five model stars generated with MESA, and discussed the physical and computational parameters used in successful runs.
We have demonstrated that \m2h can convert radial profiles to particle distributions with fidelity across 18 dex in density by applying \m2h to stellar models which vary in structure from compact, solar-mass white dwarfs to super massive, radially extended red giants. 
{We have verified \m2h's utility as an IC-generating tool using the Phantom smoothed-particle hydrodynamics code.

We intend for this tool to be openly distributed, freely available, and regularly maintained, and we anticipate that it will help in the search for solutions to a variety of problems in stellar astrophysics.}

\acknowledgements
{This work was supported by Martin Asplund, the Research School of Astronomy and Astrophysics at the Australian National University, and funding from Australian Research Council grant number DP150100250. This work was additionally supported by Brian Chaboyer and grant AST-1211384 from the American National Science Foundation.

M. Joyce would like to extend particular gratitude to Phil Taylor and Zhengwei Liu for their efforts towards validating \m2h hydrodynamically. 
M. Joyce would also like to thank Orsola de Marco for invaluable guidance on the use of Phantom for stellar applications and the invitation to Macquarie University. 
M. Joyce also gratefully acknowledges invitations to and support at Monash University via Amanda Karakas and the data visualization lab at the University of Cape Town via Thomas Jarrett, at which some of this research was conducted.
M. Joyce recognizes John Bourke for useful discussions regarding numerical methods, and Frank Timmes and Maurizio Salaris for insights on 1-D stellar models of AGB stars and white dwarfs, respectively.

D. Price acknowledges funding from the Australian Research Council via FT130100034.
}

S. Mohamed gratefully acknowledges the receipt of research funding from the National Research Foundation (NRF) of South Africa.
This work was supported in part by computational resources from the South African Astronomical Observatory and workspace accommodation from the South African Astronomical Observatory and the University of Cape Town.

\pagebreak

\bibliographystyle{apj}
\bibliography{M2H_accepted}

\end{document}